\documentclass[%
reprint,
superscriptaddress,
amsmath,
amssymb,
aps,
prl,
floatfix,
]{revtex4-2}

\usepackage{graphicx}
\usepackage{dcolumn}
\usepackage{bm}
\usepackage{hyperref}
\usepackage[dvipsnames]{xcolor}
\usepackage[mathlines]{lineno}

\begin{document}

\preprint{APS/123-QED}

\title{Non-Maxwellianity of Ion Velocity Distributions in the Earth's Magnetosheath}

\author{Louis Richard}
\email{louis.richard@irfu.se}
\affiliation{
    Swedish Institute of Space Physics, Uppsala 751 21, Sweden
}
\affiliation{
    Department of Earth, Planetary, and Space Sciences, University of California, Los Angeles, California 90095, USA
}

\author{Sergio Servidio}
\affiliation{
    Universit\`a della Calabria, Arcavacata di Rende, 87036, IT
}

\author{Ida Svenningsson}
\affiliation{
    Swedish Institute of Space Physics, Uppsala 751 21, Sweden
}
\affiliation{
    Department of Physics and Astronomy, Uppsala University, Uppsala 751 20, Sweden
}

\author{Anton V. Artemyev}
\affiliation{
    Department of Earth, Planetary, and Space Sciences, University of California, Los Angeles, California 90095, USA
}

\author{Kristopher G. Klein}
\affiliation{
    Lunar and Planetary Laboratory, University of Arizona, Tucson, AZ, USA
}

\author{Emiliya Yordanova}
\affiliation{
    Swedish Institute of Space Physics, Uppsala 751 21, Sweden
}

\author{Alexandros Chasapis}
\affiliation{
    Laboratory for Atmospheric and Space Physics, University of Colorado, Boulder, Colorado 80303, USA
}

\author{Oreste Pezzi}
\affiliation{
Istituto per la Scienza e Tecnologia dei Plasmi, Consiglio Nazionale delle Ricerche (ISTP-CNR), 70126 Bari, Italy
}

\author{Yuri V. Khotyaintsev}
\affiliation{
    Swedish Institute of Space Physics, Uppsala 751 21, Sweden
}
\affiliation{
    Department of Physics and Astronomy, Uppsala University, Uppsala 751 20, Sweden
}

\date{\today}
 
\begin{abstract}
We analyze the deviations from local thermodynamic equilibrium (LTE) of the ion velocity distribution function (iVDF) in collisionless plasma turbulence. 
Using data from the Magnetospheric Multiscale (MMS) mission, we examine the non-Maxwellianity of 439,685 iVDFs in the Earth's magnetosheath. 
We find that the iVDFs' anisotropies and the high-order non-bi-Maxwellian features are widespread and can be significant. 
Our results show that the complexity of the iVDFs is strongly influenced by the ion plasma beta and turbulence intensity, with high-order non-LTE features emerging in the presence of large-amplitude magnetic field fluctuations. 
Furthermore, our analysis indicates that turbulence-driven magnetic curvature contributes to the isotropization of the iVDFs by scattering the ions, emphasizing the complex interaction between turbulence and the velocity distribution of charged particles in collisionless plasmas.
\end{abstract}
    
\maketitle

\emph{Introduction} - 
Coulomb collisions are nearly negligible in many space and astrophysical plasma environments, e.g., accretion disks, the intracluster medium, and throughout the heliosphere~\cite{schekochihin_astrophysical_2009,verscharen_multiscale_2019}, allowing charged particle velocity distributions (VDFs) to deviate from the local thermodynamic equilibrium (LTE) Maxwell-Boltzmann distribution~\cite{krall_principles_1973}. Non-Maxwellian ion VDFs (iVDFs) are characteristic of collisionless plasma processes, including magnetic reconnection~\cite{zenitani_kinetic_2013,richard_fast_2023}, turbulence~\cite{servidio_local_2012,greco_inhomogeneous_2012,perri_deviation_2020,zhdankin_generalized_2022}, Kelvin-Helmholtz instabilities~\cite{sorriso-valvo_turbulencedriven_2019}, and shocks~\cite{parks_entropy_2012,agapitov_energy_2023}. 
\textit{In situ} observations reveal complex non-Maxwellian iVDFs features, such as anisotropies, agyrotropies, beams, flattops, and hammerheads~\cite{bale_magnetic_2009,maruca_mms_2018,martinovic_enhancement_2020,verniero_strong_2022}, providing free energy for kinetic instabilities that relax the system toward a marginally stable equilibrium through, e.g., wave-particle resonant diffusion~\cite{krall_principles_1973}. 
Recent theory suggests that the equilibrium VDFs can be universally non-Maxwellian~\cite{ewart_collisionless_2022,ewart_nonthermal_2023,ewart_relaxation_2025}. To assess the importance of these non-LTE effects, it is crucial to evaluate the occurrence of non-Maxwellian iVDFs in real collisionless plasmas. However, this has been investigated only for a limited number of case studies to date, demanding a statistical evaluation. \

\textit{In situ} observations in the solar wind and the Earth's magnetosheath suggest that iVDFs' anisotropies are constrained by the growth of the mirror, ion-cyclotron, and firehose instabilities, based on linear and quasi-linear Vlasov theory with bi-Maxwellian ions~\cite{bale_magnetic_2009,maruca_mms_2018}. 
However, the evolution of non-bi-Maxwellian iVDFs is not adequately captured by bi-Maxwellian-based theory~\cite{walters_effects_2023}. Moreover, within the framework of linear and quasi-linear Vlasov theory, the iVDFs relax toward equilibrium through resonant diffusion with weak, randomly phased perturbations~\cite{krall_principles_1973,stix_waves_1992}. 
This assumption may be violated in the presence of large-amplitude turbulence, such as the Earth's magnetosheath~\cite{maron_simulations_2001,matthaeus_nonlinear_2014} or large-amplitude quasi-monochromatic waves~\cite{li_anomalous_2022}. 
Furthermore, other mechanisms, such as pitch-angle scattering in Alfv\'enic current filaments, flow vortices, and sharp magnetic field bends, may supplement or even surpass resonant diffusion in isotropizing iVDFs~\cite{chaston_turbulent_2020,chaston_ion_2021,artemyev_superfast_2020,richard_fast_2023,lemoine_particle_2023}. 
Therefore, it is unclear how generally applicable the description of plasma relaxation provided by the bi-Maxwellian-based theory is.

To advance our understanding of the fundamental processes relaxing iVDFs and heating collisionless plasmas, we need statistical, quantitative insights into the complexity of the iVDFs and the coupling to magnetic field topology, which we provide in this study.\\

\emph{Data} - We analyze iVDFs from the Magnetospheric Multiscale (MMS) mission, investigating non-Maxwellian features --- specifically temperature anisotropy, agyrotropy, and further deviations from bi-Maxwellianity --- in turbulent space plasmas. We use magnetic field data from the fluxgate magnetometer~\cite{russell_magnetospheric_2016}, and iVDFs and their moments from the fast plasma investigation dual ion spectrometer (FPI-DIS)~\cite{pollock_fast_2016}, corrected for penetrating radiation~\cite{gershman_systematic_2019}. Data from the four spacecraft are averaged to improve statistics and reduce uncertainties. We focus on a subset of the dataset from Ref~\cite{svenningsson_classifying_2025} in the Earth's magnetosheath, a region of shocked, turbulent plasma where reconnection is frequent~\cite{retino_situ_2007,yordanova_electron_2016,voros_mms_2017,stawarz_turbulencedriven_2022}. The analysis is restricted to intervals with densities $n_i \leq 70~\mathrm{cm}^{-3}$, avoiding count-rate saturation effects, and high-quality tetrahedra, ensuring gradient errors $\lesssim 10\%$~\cite{robert_accuracy_1998}. Spacecraft separations must be smaller than the ion inertial length $d_i = \sqrt{m_i / \mu_0 n_i e^2}$, the ion gyroradius $\rho_i = \sqrt{\beta_i} d_i$ (with $\beta_i = 2\mu_0 n_i k_B T_i / B^2$), and the magnetic curvature radius $r_c = 1 / |\bm{K}|$, where $\bm{K} = \hat{\bm{b}} \cdot \nabla \hat{\bm{b}}$ is computed using the curlometer technique~\cite{chanteur_spatial_1998,chanteur_spatial_1998a}. This yields 439,685 iVDFs at 150 ms cadence ($\simeq 0.04 f_{ci}^{-1}$, with $f_{ci} = eB/2\pi m_i$).\\

\begin{figure}[!b]
    \centering
    \includegraphics[width=\linewidth]{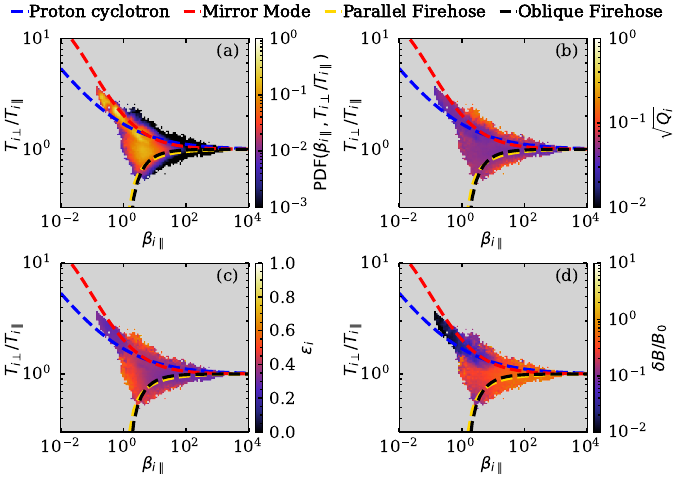}
    \caption{
        \label{fig:brazil-plots}
        Non-Maxwellianity of iVDFs in the MFA frame.  
        (a) Joint PDF of $T_{i\bot}/T_{i\parallel}$ and $\beta_{i\parallel}$.  
        (b)–(d) Conditional averages of the agyrotropy $\sqrt{Q_i}$, non-bi-Maxwellianity $\varepsilon_i$, and magnetic fluctuation amplitude $\delta B/B_0$ at $k\rho_i \simeq 1$.  
        Dashed lines mark instability thresholds for $\gamma/\omega_{ci} = 10^{-2}$ from Ref.~\cite{verscharen_collisionless_2016}.  
        Bins with fewer than nine counts are omitted due to Poisson errors~\cite{barlow_statistics_1989}.
    }
\end{figure}

\emph{Results} - We start by analyzing the non-Maxwellianity of the iVDFs in the magnetic field-aligned (MFA) frame~\cite{swisdak_quantifying_2016}. In this frame, one axis is aligned with the local magnetic field $\hat{\bm{b}} = \bm{B}/B$, and the perpendicular directions are set to equalize the remaining diagonal components of the ion temperature tensor $\mathbf{T}_i = T_{i\parallel} \hat{\bm{b}}\hat{\bm{b}} + T_{i\bot} (\mathbf{1} - \hat{\bm{b}}\hat{\bm{b}}) + \bm{\Theta}$, where $T_{i\parallel}$ and $T_{i\bot}$ are the parallel and perpendicular temperatures, and $\bm{\Theta}$ is the agyrotropic part with zero diagonal. Fig.~\ref{fig:brazil-plots}(a) shows the joint probability density function (PDF) in the $(\beta_{i\parallel}, T_{i\bot}/T_{i\parallel})$ parameter space. While this approach is commonly adopted, we will later show its limitations. We find that the majority (68\%) of the iVDFs lie within the stability limits predicted by linear Vlasov theory. 
In particular, the perpendicular temperature anisotropy $T_{i\bot}/T_{i\parallel} > 1$ is well constrained by the mirror-mode instability threshold, consistent with previous observations~\cite{maruca_mms_2018}. 
However, 32\% of iVDFs fall outside these stability limits, indicating that unstable distributions are commonly observed and/or that the theoretical boundaries are inaccurate due to deviations from the bi-Maxwellian model VDF.

To quantify the non-bi-Maxwellianity, we now examine the agyrotropy of the iVDFs. 
We use Swisdak's agyrotropy parameter, defined as~\cite{swisdak_quantifying_2016}:

\begin{equation}
    \label{eq:swisdak}
    \sqrt{Q_i} = \sqrt{\frac{T_{i12}^2 + T_{i13}^2 + T_{i23}^2}{T_{i\bot}^2 + 2T_{i\bot} T_{i\parallel}}},
\end{equation}
\noindent
where $\sqrt{Q_i}\in [0, 1]$, with $\sqrt{Q_i}=0$ for a perfectly gyrotropic iVDF. 
Here, $T_{i12}$, $T_{i13}$, and $T_{i23}$ are the off-diagonal components of $\bm{\Theta}$. We find that the agyrotropy is generally very small, with $\sqrt{Q_i} = 0.04^{+0.02}_{-0.02}$ [Fig.~\ref{fig:brazil-plots}(b)]. 
However, $4\%$ of the iVDFs exhibit significant agyrotropy with $\sqrt{Q_i} \gtrsim 0.1$, which is typically observed in current sheets and magnetic reconnection regions~\cite{motoba_new_2022}. 
This result indicates that the iVDFs are overall gyrotropic, with some extreme agyrotropy appearing in intervals exceeding stability conditions.

Complex, gyrotropic yet non-bi-Maxwellian, iVDFs such as beams, kappa, and flattops are ubiquitous in collisionless plasma~\cite{martinovic_enhancement_2020, verniero_strong_2022,ewart_collisionless_2022,ewart_nonthermal_2023,ewart_relaxation_2025}. To quantify their non-bi-Maxwellianity, we adopt the normalized $L^1$ norm of the non-bi-Maxwellian phase-space density residue $\delta f_i = f_i - f_{ibM}$, defined in Ref.~\cite{graham_nonmaxwellianity_2021}:

\begin{equation}
    \label{eq:graham}
    \varepsilon_i = \frac{1}{2 n_i} \iiint |\delta f_i(v, \theta, \phi)| v^2 \sin \theta \mathrm{d} v \mathrm{d}\theta \mathrm{d}\phi,
\end{equation}

\noindent
where $v$, $\theta$, and $\phi$ are sampled in the FPI-DIS instrument grid. 
Here, $f_i$ is the measured iVDF, and $f_{i\mathrm{bM}}$ is the bi-Maxwellian with the same moments. 
The prefactor $1/2n_i$ ensures that $\varepsilon_i = 0$ for a perfect bi-Maxwellian $f_i$, and $\varepsilon_i=1$ for maximal deviation.
The integral is evaluated summing up FPI-DIS bins within $v_{min} \leq v \leq v_{max}$~\cite{graham_nonmaxwellianity_2021,richard_fast_2023} where $f_i$ is at least $1\sigma$ above the penetrating radiation model~\cite{gershman_systematic_2019} and the noise floor (one count level), assuming that other sources of error are negligible. 
Typically, $v_{min} \simeq 0.5 v_{thi}$ and $v_{max} \simeq 6.8 v_{thi}$, where $v_{thi}=\sqrt{2k_B T_i/m_i}$ denotes the ion thermal speed. 
We find an average non-bi-Maxwellianity of $\varepsilon_i = 0.36^{+0.11}_{-0.07}$ [Fig.~\ref{fig:brazil-plots}(c)], comparable to values reported in reconnection outflows ($\varepsilon_i =0.39^{+0.08}_{-0.07}$)~\cite{richard_fast_2023}.
We note that the magnitude of $\varepsilon_i$ may be offset due to the finite grid resolution and limited energy range systematic errors~\cite{graham_nonmaxwellianity_2021,richard_fast_2023}. 
The non-bi-Maxwellianity has minima near marginal stability ($\varepsilon_i \simeq 0.29$). Larger values reaching $\simeq 0.90$ emerge on both sides of the theoretical stability boundaries. 
This indicates that iVDFs can have substantial high-order non-bi-Maxwellian features even at moderate anisotropies.

We now investigate magnetic field fluctuations associated with iVDFs, focusing on ion-scale fluctuations at $k\rho_i \simeq 1$. 
We assume a power-law decay in the power spectral density and apply Taylor's hypothesis, which holds since the ion bulk speed $V_i$ satisfies $V_i \simeq 4 V_A > V_A$. Here, $V_A = B_0 / \sqrt{\mu_0 n_i m_i}$ is the Alfv\'en speed, and $B_0 = |\langle \bm{B} \rangle_{30\mathrm{s}}|$ is the background magnetic field, where $\langle \cdot \rangle_{30\mathrm{s}}$ represents a 30-second average. We choose the averaging window to span several ($\sim 3 -10$) correlation lengths $\lambda_c$~\cite{stawarz_turbulencedriven_2022,svenningsson_classifying_2025}. We compute the normalized magnetic field wave amplitude $\delta B / B_0$, using a high-pass filter with a cutoff frequency corresponding to $k\rho_i \simeq 1$, meaning $f \geq \langle V_i \rangle_{30\textrm{s}} / 2\pi \langle \rho_i \rangle_{30\textrm{s}}$. For $\beta_{i\parallel} < 1$ and $T_{i\bot} / T_{i\parallel} > 1$, near marginal stability [Fig.~\ref{fig:brazil-plots}(a)], we find that the magnetic fluctuation amplitude is $\delta B / B_0 \sim 10^{-2}$ [Fig.~\ref{fig:brazil-plots}(d)]. 
In contrast, for $\beta_{i\parallel} > 1$ and $T_{i\bot} / T_{i\parallel} < 1$, ion-scale magnetic fluctuations intensify, reaching $\delta B / B_0 \simeq 0.44$ near instability thresholds and regions of large non-bi-Maxwellianity $\varepsilon_i$ [Fig.~\ref{fig:brazil-plots}(c)]. 
This suggests that marginally stable and non-bi-Maxwellian iVDFs are associated with large amplitude ion scale magnetic field fluctuations.

\begin{figure}[!t]
    \centering
    \includegraphics[width=\linewidth]{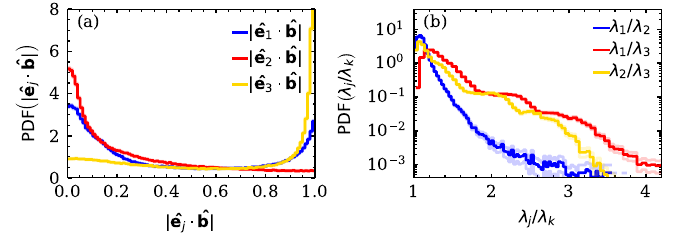}
    \caption{
        \label{fig:pca}
        Non-Maxwellianity of iVDF in the Minimum Variance Frame. 
        (a) PDF of the principal directions $\hat{\bm{e}}_i$ relative to the local magnetic field $\hat{\bm{b}}$. 
        (c) PDF of the eigenvalue ratios $\lambda_1/\lambda_2$, $\lambda_1/\lambda_3$, and $\lambda_2/\lambda_3$. 
        Shaded regions represent standard deviations assuming Poisson-distributed statistics~\cite{barlow_statistics_1989}.
    }
\end{figure}

To determine if the magnetic field direction $\hat{\bm{b}}$ is a preferred direction of the iVDFs, we investigate their orientation with respect to the MFA frame.
We determine the minimum variance frame (MVF) of the iVDFs by calculating the eigenvalues $\left\{\lambda_j\right\}_{j\in [1, 3]}$ and the corresponding principal directions $\left \{ \hat{\bm{e}}_j\right\}_{j\in [1, 3]}$ of $\bm{T}_i$ such that $\lambda_1 \geq \lambda_2 \geq \lambda_3$~\cite{servidio_local_2012,servidio_kinetic_2015,richard_fast_2023}. 
We note that in the special case where $\bm{\hat{e}}_1 = \bm{\hat{b}}$ and $\lambda_2 = \lambda_3=T_{i\bot}$, i.e., gyrotropic ($\bm{\Theta}_i=\bm{0}$), the MVF and MFA frames coincide.
We find that the maximum variance direction $\hat{\bm{e}}_1$ is predominantly perpendicular to $\hat{\bm{b}}$ [Fig.~\ref{fig:pca}(a)]. 
Meanwhile, the minimum variance direction $\hat{\bm{e}}_3$ generally aligns with $\hat{\bm{b}}$. 
Additionally, the eigenvalue ratios are $\lambda_1 / \lambda_2 = 1.10^{+0.09}_{-0.05}$ and $\lambda_1 / \lambda_3 =1.30^{+0.32}_{-0.14}$ [Fig.~\ref{fig:pca}(b)]. 
Given the orientation of $\hat{\bm{e}}_1$ and $\hat{\bm{e}}_3$ relative to $\hat{\bm{b}}$, $\lambda_1 / \lambda_2$ indicates agyrotropy, while $\lambda_1 / \lambda_3$ represents anisotropy. 
The $\lambda_1 / \lambda_2$ distribution exhibits a distinct non-Gaussian power-law tail with a kurtosis $\mu_4/\mu_2^2 \approx 74$. 
However, in some cases, $\hat{\bm{e}}_1$ closely aligns with $\hat{\bm{b}}$ ($|\hat{\bm{e}}_1 \cdot \hat{\bm{b}}| > 0.8$), coinciding with $\lambda_1 / \lambda_2 \simeq 1.15^{+0.15}_{-0.08}$ and $\lambda_1 / \lambda_3 \simeq 1.26^{+0.19}_{-0.11}$ (not shown), suggesting nearly isotropic iVDFs with $\lambda_1 \sim \lambda_2 \sim \lambda_3$. 
This confirms a broad anisotropy distribution, predominantly with $T_{i\bot} > T_{i\parallel}$, while agyrotropy remains generally small but can reach significant levels.

\begin{figure}[!b]
    \centering
    \includegraphics[width=\linewidth]{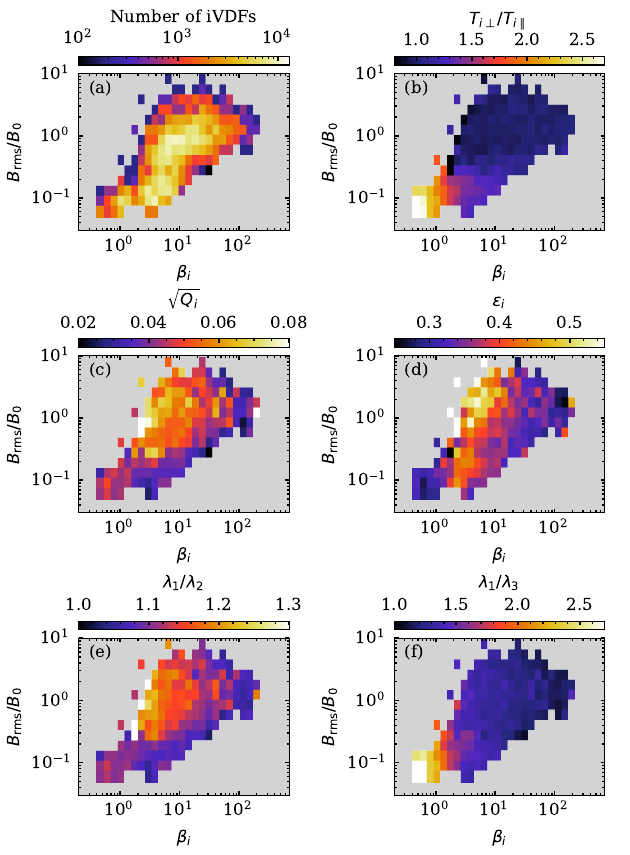}
    \caption{
        \label{fig:turbulence}
        Conditional averages of the non-Maxwellianity of the iVDFs in the $(\beta_i, B_{\mathrm{rms}}/B_0)$ space. 
        (a) Number of iVDFs. 
        (b) Temperature anisotropy $T_{i\bot}/T_{i\parallel}$ 
        (c) Agyrotropy $\sqrt{Q_i}$. 
        (d) Non-bi-Maxwellianity $\varepsilon_i$. 
        (e) Maximum to intermediate $\lambda_1/\lambda_2$ and (f) maximum to minimum $\lambda_1/\lambda_3$ eigenvalue ratios, respectively.
    }
\end{figure}

Thus far, we have analyzed the non-Maxwellianity of the iVDFs without considering turbulence effects. 
However, numerical simulations suggest that turbulence levels and $\beta_i$ significantly influence deviations from LTE~\cite{servidio_proton_2014}. 
To explore this dependence, we bin our dataset based on $\beta_i$ and the turbulence level, quantified by the variability of the magnetic field, $B_{\mathrm{rms}}/B_0$ [Fig.~\ref{fig:turbulence}(a)], where $B_{\mathrm{rms}} = \sqrt{\langle | \bm{B} - \bm{B}_0|^2 \rangle_{30\mathrm{s}}}$ is the root mean square magnetic fluctuations with $\bm{B}_0 = \langle \bm{B}\rangle_{30\mathrm{s}}$ the background magnetic field. 
We then compute the conditional average of the non-Maxwellianity measures $T_{i\bot}/T_{i\parallel}$, $\sqrt{Q_i}$, $\varepsilon_i$, $\lambda_1/\lambda_2$, and $\lambda_1/\lambda_3$ in the ($\beta_i,~B_{\mathrm{rms}}/B_0$) parameter space based on the averaged quantities within each 30-second window.

This classification reveals two distinct clusters at $(\beta_i \sim 1,~B_{\mathrm{rms}}/B_0 \sim 0.1)$ and $(\beta_i \sim 10,~B_{\mathrm{rms}}/B_0 \sim 1)$, corresponding to the quasi-perpendicular and quasi-parallel magnetosheath, respectively~\cite{svenningsson_classifying_2025}. 
We find that the temperature anisotropy, $T_{i\bot}/T_{i\parallel}$, decreases with increasing $\beta_i$ and $B_{\mathrm{rms}}/B_0$ [Fig.~\ref{fig:turbulence}(b)]. 
Due to the correlation between $\beta_i$ and $B_{\mathrm{rms}}/B_0$, and the lack of data at ($\beta_i \sim 1,~B_{\mathrm{rms}}/B_0 \sim 1$) and ($\beta_i \sim 10,~B_{\mathrm{rms}}/B_0 \sim 0.1$), it remains unclear whether this decrease is a direct effect of turbulence or an indirect consequence of its relationship with $\beta_i$.  

In contrast, the agyrotropy $\sqrt{Q_i}$ [Fig.~\ref{fig:turbulence}(c)] and the non-bi-Maxwellianity $\varepsilon_i$ decrease with increasing $\beta_i$ but increase with $B_{\mathrm{rms}}/B_0$ [Fig.~\ref{fig:turbulence}(c),~\ref{fig:turbulence}(d)]. 
Notably, at $\beta_i\sim 5,~B_{\mathrm{rms}}/B_0 \sim 1$, we find values reaching $\sqrt{Q_i}\simeq 0.07$ and $\varepsilon_i \simeq 0.5$, which significantly exceed the dataset mean ($\sqrt{Q_i} = 0.04^{+0.02}_{-0.02}$, $\varepsilon_i = 0.36^{+0.11}_{-0.07}$) [Figs.~\ref{fig:brazil-plots}(b),~\ref{fig:brazil-plots}(c)]. 
Furthermore, the eigenvalue ratios $\lambda_1/\lambda_2$ and $\lambda_1/\lambda_3$ closely follow the trends of agyrotropy and anisotropy, respectively [Figs.~\ref{fig:turbulence}(b),~\ref{fig:turbulence}(c),~\ref{fig:turbulence}(e),~\ref{fig:turbulence}(f)]. 
This indicates that in the low-$\beta_i$, weakly turbulent regime, iVDFs are compressed along the magnetic field, i.e., $\lambda_1\sim \lambda_2 > \lambda_3$, with little high-order non-bi-Maxwellianity. 
In contrast, in the high-$\beta_i$, strongly turbulent regime, the iVDFs become globally isotropic with pronounced high-order non-bi-Maxwellian features.

\begin{figure}[!t]
    \centering
    \includegraphics[width=\linewidth]{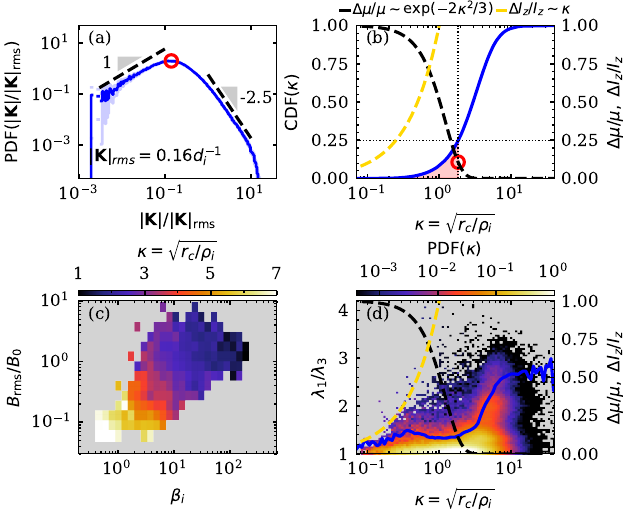}
    \caption{
        \label{fig:curvature}
        Magnetic field curvature. 
        (a) PDF of the normalized magnetic field curvature $|\bm{K}|/|\bm{K}|_{\mathrm{rms}}$, with black dashed lines showing predicted slopes from Ref.~\cite{bandyopadhyay_situ_2020a}. 
        Shaded regions represent standard deviations assuming Poisson-distributed statistics~\cite{barlow_statistics_1989}. 
        (b) Cumulative distribution function (CDF) of the adiabaticity parameter $\kappa$. 
        (c) Conditional averages of $\kappa$ in the $(\beta_i, B_{\mathrm{rms}}/B_0)$ space. 
        (d) Joint PDF of the eigenvalue ratio $\lambda_1/\lambda_3$ and $\kappa$, with the blue line marking the 95th percentile. 
        The black and gold lines show theoretical jumps $\Delta \mu / \mu$ and $\Delta I_z / I_z$, respectively~\cite{buchner_regular_1989}.
    }
\end{figure}

To understand how iVDFs relax toward LTE, we examine the magnetic field curvature relative to the ion gyroradius. 
Indeed, in addition to wave-particle interactions, sharp magnetic field bends can efficiently scatter ions~\cite{buchner_regular_1989,artemyev_superfast_2020,richard_fast_2023,lemoine_particle_2023}.  
The PDF of the normalized magnetic curvature $|\bm{K}| / |\bm{K}|_{\mathrm{rms}}$ resembles a double Pareto lognormal distribution [Fig.~\ref{fig:curvature}a], supporting numerical simulations and earlier observations~\cite{yang_role_2019,bandyopadhyay_situ_2020a}.
Notably, the PDF peaks at $|\bm{K}|^\star \simeq 0.14 |\bm{K}|_{\mathrm{rms}} \simeq 0.02 d_i^{-1}$ (red circle), which corresponds to $|\bm{K}|^\star \sim \lambda_c^{-1}$, with $\lambda_c\sim 10 - 100 d_i$~\cite{stawarz_turbulencedriven_2022}. 
Moreover, for $|\bm{K}| \ll |\bm{K}|^\star$ and $|\bm{K}| \gg |\bm{K}|^\star$, the PDF approximately follows power-law slopes of $1$ and $-2.5$, respectively. Theoretical predictions suggest that these asymptotic limits result from the large curvature $\bm{K}$ where the magnetic field strength is weak, and the low curvature associated with small magnetic tension forces, respectively~\cite{yang_role_2019,bandyopadhyay_situ_2020a}.

We compute the adiabaticity parameter, $\kappa = \sqrt{r_c / \rho_i}$, where $r_c = 1 / |\bm{K}|$ is the radius of curvature of the magnetic field, as a proxy for ion motion dynamics~\cite{buchner_regular_1989} [Fig.~\ref{fig:curvature}b]. 
For $\kappa \gg 1$, ion motion is adiabatic, with the magnetic moment $\mu$ conserved to an accuracy of $\Delta \mu / \mu \sim \exp(-2\kappa^2 / 3)$~\cite{birmingham_pitch_1984,buchner_regular_1989}. 
At $\kappa \sim 1$, resonance occurs, leading to chaotic ion motion and strong scattering, characterized by $\Delta \mu / \mu \sim 1$~\cite{birmingham_pitch_1984,buchner_regular_1989}. 
For $\kappa \ll 1$, the motion is quasi-adiabatic, with the generalized magnetic moment $ I_z $ approximately conserved as $\Delta I_z / I_z \sim \kappa$ for compressional and $\sim 1$ for force-free discontinuities~\cite{burkhart_particle_1995,artemyev_rapid_2014}. 
We find that for 25\% of the iVDFs $\kappa \leq 1.8$ (red shaded area), corresponding to $\Delta \mu / \mu \geq 0.1$ (red circle). 
Since $\kappa$ is estimated using the thermal ion speed $v_{thi\bot}$, ions with $v_{i\bot} > v_{thi\bot}$ may remain non-adiabatic even when $\kappa \gg 1$. 
This indicates that a substantial fraction of the ions follow non-adiabatic orbits, thereby breaking the conservation of magnetic moment.

The adiabaticity parameter $\kappa$ decreases with increasing $\beta_i$ and $B_{\mathrm{rms}} / B_0$ due to its temperature and magnetic field strength dependence [Fig.~\ref{fig:curvature}c]. 
Notably, its variation with $\beta_i$ and $B_{\mathrm{rms}} / B_0$ closely follows the trend of $T_{i\bot}/T_{i\parallel}$ [Figs.~\ref{fig:turbulence}(b),~\ref{fig:turbulence}(f)]. 
At $(\beta_i \sim 1,~B_{\mathrm{rms}} / B_0 \sim 0.1)$, we find $\kappa \simeq 7$, corresponding to $\Delta \mu / \mu \simeq 2 \times 10^{-16}$, indicating negligible curvature scattering. 
In contrast, at $(\beta_i \sim 100,~B_{\mathrm{rms}} / B_0 \sim 1)$, $\kappa \sim 1$, leading to $\Delta \mu / \mu \simeq 1$, and ions experience strong scattering. 
Moreover, $\kappa$ is strongly correlated with $\lambda_1/\lambda_3$ [Fig.~\ref{fig:curvature}d], whose 95th percentile reaches a minimum at $\kappa \simeq 1.21$, corresponding to $\Delta \mu / \mu \simeq 0.38$, near the expected maximum scattering at $\kappa = 1$. 
This suggests that the isotropization of iVDFs results from the violation of adiabatic invariance due to sharp magnetic field bends.\\

\emph{Discussion} - 
Our analysis shows that iVDFs in Earth's magnetosheath can deviate significantly from Maxwell-Boltzmann LTE. 
Many iVDFs are predicted to be unstable to temperature anisotropy-driven instabilities. 
However, both stable and unstable iVDFs also display notable high-order non-bi-Maxwellian features, suggesting that the bi-Maxwellian model commonly used in linear Vlasov theory may be insufficient to assess the iVDFs' stability~\cite{klein_majority_2018,martinovic_iondriven_2021,martinovic_iondriven_2023,walters_effects_2023}. We note that we do not account for contributions from other species, such as alpha particles, which may influence the non-Maxwellianity and the stability limits~\cite{maruca_instabilitydriven_2012,chen_multispecies_2016,martinovic_iondriven_2021,deweese_alpha_2022}.
Non-LTE iVDFs are often associated with large amplitude magnetic field fluctuations. These fluctuations may be ion scale magnetic shears, pushing the iVDFs further from LTE rather than relaxing them toward marginal stability, as one might expect from electromagnetic waves~\cite{osman_kinetic_2012,servidio_proton_2014,kunz_firehose_2014,delsarto_pressure_2016}. 
Alternatively, these unstable iVDFs may originate at the bow shock and persist due to the plasma's slow transit time across the magnetosheath compared with the relaxation timescales~\cite{richard_fast_2023}.

We demonstrate that the non-Maxwellianity of iVDFs depends on $\beta_i$ and the turbulence level $B_{\mathrm{rms}}/B_0$. 
As $B_{\mathrm{rms}}/B_0$ increases, high-order non-Maxwellian features, including agyrotropy, become more pronounced, suggesting that these are driven by the turbulence. 
Numerical simulations and \textit{in situ} observations suggest that turbulence-generated current sheets create such features~\cite{greco_inhomogeneous_2012,servidio_local_2012,perri_deviation_2020}. 
Additionally, the agyrotropy follows a non-Gaussian power-law distribution, mirroring the behavior of magnetic field increments in turbulence~\cite{sundkvist_dissipation_2007,yordanova_magnetosheath_2008}. This suggests turbulence may be linked to these high-order non-LTE features. 
Moreover, the measured fine-scale velocity-space features may be tied to a concurrent phase-space cascade~\cite{schekochihin_astrophysical_2009,tatsuno_nonlinear_2009,servidio_magnetospheric_2017}, further highlighting the complex interplay between turbulence and iVDF evolution.

Interestingly, in weakly magnetized, highly turbulent regime $(\beta_i \gg 1,~B_{\mathrm{rms}}/B_0 \sim 1)$ iVDFs are globally isotropic despite high-order non-Maxwellian features. We note that while the iVDF is globally isotropic, it may contain anisotropic parts balancing each other.
Nevertheless, in these conditions, magnetic curvature induced by turbulence can be of the order of the ion gyroradius scale or smaller. 
As a result, ions interact with the turbulence-generated curved magnetic fields, undergoing pitch-angle scattering~\cite{birmingham_pitch_1984,buchner_regular_1989, artemyev_charged_2016, artemyev_superfast_2020,malara_chargedparticle_2021}. 
Numerical simulations further suggest that such strongly curved magnetic fields enhance particle transport and diffusion~\cite{servidio_explosive_2016, lemoine_firstprinciples_2022, lemoine_particle_2023}. 
This indicates that turbulence distorts magnetic fields across all scales down to sub-ion scales, leading to ion phase-space diffusion and energization through interaction with these curved fields.\\

\emph{Conclusion} - We present the first quantitative, statistical analysis of iVDFs in the turbulent, collisionless plasma of Earth's magnetosheath. Using a framework to quantify deviations from LTE, we show that departures from a bi-Maxwellian are both widespread and significant. Our novel parametric analysis reveals that these deviations decrease with $\beta_i$ and increase with the turbulence level. The ubiquity of such non-Maxwellian features raises a question about the degree to which the bi-Maxwellian-based theory is applicable in describing the relaxation of the shocked turbulent plasma. \

We demonstrate that curvature scattering in the presence of large-amplitude turbulence contributes to the relaxation of non-LTE iVDFs, pointing to alternative, turbulence-driven relaxation pathways. These findings advance our understanding of the kinetic response of ions to turbulence and highlight the importance of turbulence-driven processes in shaping the velocity space dynamics of space and astrophysical plasmas.\\

\begin{acknowledgments}
\emph{Acknowledgments} - 
We thank the MMS team for data access and support. L.R. thanks D. J. Gershman for helpful suggestions. L.R. acknowledges support from the Knut and Alice Wallenberg Foundation (Dnr. 2022.0087), the Swedish National Space Agency grant 192/20, and the Royal Swedish Academy of Sciences grant AST2024-0015. 
S.S. acknowledges the Space It Up project funded by the Italian Space Agency, ASI, and the Ministry of University and Research, MUR, under contract n. 2024-5-E.0 - CUP n. I53D24000060005.
I.S. acknowledges support from the Swedish Research Council Grant 2016‐0550 and the Swedish National Space Agency Grant 158/16.
K. G. K. was supported by NASA ECIP Grant
No. 80NSSC19K0912.
A.C. acknowledges support from NASA Grants 80NSSC21K0454, 80NSSC22K0688, and 80NSSC24K0172.
O.P. acknowledges the project "2022KL38BK - The ULtimate fate of TuRbulence from space to laboratory plAsmas (ULTRA)" (Master CUP B53D23004850006), by the Italian Ministry of University and Research, funded under the National Recovery and Resilience Plan (NRRP), Mission 4 – Component C2 – Investment 1.1, "Fondo per il Programma Nazionale di Ricerca e Progetti di Rilevante Interesse Nazionale (PRIN 2022)" (PE9) by the European Union – NextGenerationEU.
This research was supported by the International Space Science Institute (ISSI) in Bern through the ISSI International Team project \#23-588 ("Unveiling Energy Conversion and Dissipation in Non-Equilibrium Space Plasmas").
Data analysis used the \verb+pyrfu+ analysis package \footnote{See \url{https://pypi.org/project/pyrfu/}.}.
\end{acknowledgments}

\bibliographystyle{apsrev4-2}
\bibliography{main}

\begin{thebibliography}{70}%
\makeatletter
\providecommand \@ifxundefined [1]{%
 \@ifx{#1\undefined}
}%
\providecommand \@ifnum [1]{%
 \ifnum #1\expandafter \@firstoftwo
 \else \expandafter \@secondoftwo
 \fi
}%
\providecommand \@ifx [1]{%
 \ifx #1\expandafter \@firstoftwo
 \else \expandafter \@secondoftwo
 \fi
}%
\providecommand \natexlab [1]{#1}%
\providecommand \enquote  [1]{``#1''}%
\providecommand \bibnamefont  [1]{#1}%
\providecommand \bibfnamefont [1]{#1}%
\providecommand \citenamefont [1]{#1}%
\providecommand \href@noop [0]{\@secondoftwo}%
\providecommand \href [0]{\begingroup \@sanitize@url \@href}%
\providecommand \@href[1]{\@@startlink{#1}\@@href}%
\providecommand \@@href[1]{\endgroup#1\@@endlink}%
\providecommand \@sanitize@url [0]{\catcode `\\12\catcode `\$12\catcode `\&12\catcode `\#12\catcode `\^12\catcode `\_12\catcode `\%12\relax}%
\providecommand \@@startlink[1]{}%
\providecommand \@@endlink[0]{}%
\providecommand \url  [0]{\begingroup\@sanitize@url \@url }%
\providecommand \@url [1]{\endgroup\@href {#1}{\urlprefix }}%
\providecommand \urlprefix  [0]{URL }%
\providecommand \Eprint [0]{\href }%
\providecommand \doibase [0]{https://doi.org/}%
\providecommand \selectlanguage [0]{\@gobble}%
\providecommand \bibinfo  [0]{\@secondoftwo}%
\providecommand \bibfield  [0]{\@secondoftwo}%
\providecommand \translation [1]{[#1]}%
\providecommand \BibitemOpen [0]{}%
\providecommand \bibitemStop [0]{}%
\providecommand \bibitemNoStop [0]{.\EOS\space}%
\providecommand \EOS [0]{\spacefactor3000\relax}%
\providecommand \BibitemShut  [1]{\csname bibitem#1\endcsname}%
\let\auto@bib@innerbib\@empty
\bibitem [{\citenamefont {Schekochihin}\ \emph {et~al.}(2009)\citenamefont {Schekochihin}, \citenamefont {Cowley}, \citenamefont {Dorland}, \citenamefont {Hammett}, \citenamefont {Howes}, \citenamefont {Quataert},\ and\ \citenamefont {Tatsuno}}]{schekochihin_astrophysical_2009}%
  \BibitemOpen
  \bibfield  {author} {\bibinfo {author} {\bibfnamefont {A.~A.}\ \bibnamefont {Schekochihin}}, \bibinfo {author} {\bibfnamefont {S.~C.}\ \bibnamefont {Cowley}}, \bibinfo {author} {\bibfnamefont {W.}~\bibnamefont {Dorland}}, \bibinfo {author} {\bibfnamefont {G.~W.}\ \bibnamefont {Hammett}}, \bibinfo {author} {\bibfnamefont {G.~G.}\ \bibnamefont {Howes}}, \bibinfo {author} {\bibfnamefont {E.}~\bibnamefont {Quataert}},\ and\ \bibinfo {author} {\bibfnamefont {T.}~\bibnamefont {Tatsuno}},\ }\href {https://doi.org/10.1088/0067-0049/182/1/310} {\bibfield  {journal} {\bibinfo  {journal} {Astrophys. J. Suppl. Ser.}\ }\textbf {\bibinfo {volume} {182}},\ \bibinfo {pages} {310} (\bibinfo {year} {2009})}\BibitemShut {NoStop}%
\bibitem [{\citenamefont {Verscharen}\ \emph {et~al.}(2019)\citenamefont {Verscharen}, \citenamefont {Klein},\ and\ \citenamefont {Maruca}}]{verscharen_multiscale_2019}%
  \BibitemOpen
  \bibfield  {author} {\bibinfo {author} {\bibfnamefont {D.}~\bibnamefont {Verscharen}}, \bibinfo {author} {\bibfnamefont {K.~G.}\ \bibnamefont {Klein}},\ and\ \bibinfo {author} {\bibfnamefont {B.~A.}\ \bibnamefont {Maruca}},\ }\href {https://doi.org/10.1007/s41116-019-0021-0} {\bibfield  {journal} {\bibinfo  {journal} {Living Rev. Solar Phys.}\ }\textbf {\bibinfo {volume} {16}},\ \bibinfo {pages} {5} (\bibinfo {year} {2019})}\BibitemShut {NoStop}%
\bibitem [{\citenamefont {Krall}\ and\ \citenamefont {Trivelpiece}(1973)}]{krall_principles_1973}%
  \BibitemOpen
  \bibfield  {author} {\bibinfo {author} {\bibfnamefont {N.~A.}\ \bibnamefont {Krall}}\ and\ \bibinfo {author} {\bibfnamefont {A.~W.}\ \bibnamefont {Trivelpiece}},\ }\href@noop {} {\emph {\bibinfo {title} {Principles of Plasma Physics}}}\ (\bibinfo  {publisher} {McGraw-Hill},\ \bibinfo {year} {1973})\BibitemShut {NoStop}%
\bibitem [{\citenamefont {Zenitani}\ \emph {et~al.}(2013)\citenamefont {Zenitani}, \citenamefont {Shinohara}, \citenamefont {Nagai},\ and\ \citenamefont {Wada}}]{zenitani_kinetic_2013}%
  \BibitemOpen
  \bibfield  {author} {\bibinfo {author} {\bibfnamefont {S.}~\bibnamefont {Zenitani}}, \bibinfo {author} {\bibfnamefont {I.}~\bibnamefont {Shinohara}}, \bibinfo {author} {\bibfnamefont {T.}~\bibnamefont {Nagai}},\ and\ \bibinfo {author} {\bibfnamefont {T.}~\bibnamefont {Wada}},\ }\href {https://doi.org/10.1063/1.4821963} {\bibfield  {journal} {\bibinfo  {journal} {Phys. Plasmas}\ }\textbf {\bibinfo {volume} {20}},\ \bibinfo {pages} {092120} (\bibinfo {year} {2013})}\BibitemShut {NoStop}%
\bibitem [{\citenamefont {Richard}\ \emph {et~al.}(2023)\citenamefont {Richard}, \citenamefont {Khotyaintsev}, \citenamefont {Graham}, \citenamefont {Vaivads}, \citenamefont {Gershman},\ and\ \citenamefont {Russell}}]{richard_fast_2023}%
  \BibitemOpen
  \bibfield  {author} {\bibinfo {author} {\bibfnamefont {L.}~\bibnamefont {Richard}}, \bibinfo {author} {\bibfnamefont {Y.~V.}\ \bibnamefont {Khotyaintsev}}, \bibinfo {author} {\bibfnamefont {D.~B.}\ \bibnamefont {Graham}}, \bibinfo {author} {\bibfnamefont {A.}~\bibnamefont {Vaivads}}, \bibinfo {author} {\bibfnamefont {D.~J.}\ \bibnamefont {Gershman}},\ and\ \bibinfo {author} {\bibfnamefont {C.~T.}\ \bibnamefont {Russell}},\ }\href {https://doi.org/10.1103/PhysRevLett.131.115201} {\bibfield  {journal} {\bibinfo  {journal} {Phys. Rev. Lett.}\ }\textbf {\bibinfo {volume} {131}},\ \bibinfo {pages} {115201} (\bibinfo {year} {2023})}\BibitemShut {NoStop}%
\bibitem [{\citenamefont {Servidio}\ \emph {et~al.}(2012)\citenamefont {Servidio}, \citenamefont {Valentini}, \citenamefont {Califano},\ and\ \citenamefont {Veltri}}]{servidio_local_2012}%
  \BibitemOpen
  \bibfield  {author} {\bibinfo {author} {\bibfnamefont {S.}~\bibnamefont {Servidio}}, \bibinfo {author} {\bibfnamefont {F.}~\bibnamefont {Valentini}}, \bibinfo {author} {\bibfnamefont {F.}~\bibnamefont {Califano}},\ and\ \bibinfo {author} {\bibfnamefont {P.}~\bibnamefont {Veltri}},\ }\href {https://doi.org/10.1103/PhysRevLett.108.045001} {\bibfield  {journal} {\bibinfo  {journal} {Phys. Rev. Lett.}\ }\textbf {\bibinfo {volume} {108}},\ \bibinfo {pages} {045001} (\bibinfo {year} {2012})}\BibitemShut {NoStop}%
\bibitem [{\citenamefont {Greco}\ \emph {et~al.}(2012)\citenamefont {Greco}, \citenamefont {Valentini}, \citenamefont {Servidio},\ and\ \citenamefont {Matthaeus}}]{greco_inhomogeneous_2012}%
  \BibitemOpen
  \bibfield  {author} {\bibinfo {author} {\bibfnamefont {A.}~\bibnamefont {Greco}}, \bibinfo {author} {\bibfnamefont {F.}~\bibnamefont {Valentini}}, \bibinfo {author} {\bibfnamefont {S.}~\bibnamefont {Servidio}},\ and\ \bibinfo {author} {\bibfnamefont {W.~H.}\ \bibnamefont {Matthaeus}},\ }\href {https://doi.org/10.1103/PhysRevE.86.066405} {\bibfield  {journal} {\bibinfo  {journal} {Phys. Rev. E}\ }\textbf {\bibinfo {volume} {86}},\ \bibinfo {pages} {066405} (\bibinfo {year} {2012})}\BibitemShut {NoStop}%
\bibitem [{\citenamefont {Perri}\ \emph {et~al.}(2020)\citenamefont {Perri}, \citenamefont {Perrone}, \citenamefont {Yordanova}, \citenamefont {{Sorriso-Valvo}}, \citenamefont {Paterson}, \citenamefont {Gershman}, \citenamefont {Giles}, \citenamefont {Pollock}, \citenamefont {Dorelli}, \citenamefont {Avanov}, \citenamefont {Lavraud}, \citenamefont {Saito}, \citenamefont {Nakamura}, \citenamefont {Fischer}, \citenamefont {Baumjohann}, \citenamefont {Plaschke}, \citenamefont {Narita}, \citenamefont {Magnes}, \citenamefont {Russell}, \citenamefont {Strangeway}, \citenamefont {Contel}, \citenamefont {Khotyaintsev},\ and\ \citenamefont {Valentini}}]{perri_deviation_2020}%
  \BibitemOpen
  \bibfield  {author} {\bibinfo {author} {\bibfnamefont {S.}~\bibnamefont {Perri}}, \bibinfo {author} {\bibfnamefont {D.}~\bibnamefont {Perrone}}, \bibinfo {author} {\bibfnamefont {E.}~\bibnamefont {Yordanova}}, \bibinfo {author} {\bibfnamefont {L.}~\bibnamefont {{Sorriso-Valvo}}}, \bibinfo {author} {\bibfnamefont {W.~R.}\ \bibnamefont {Paterson}}, \bibinfo {author} {\bibfnamefont {D.~J.}\ \bibnamefont {Gershman}}, \bibinfo {author} {\bibfnamefont {B.~L.}\ \bibnamefont {Giles}}, \bibinfo {author} {\bibfnamefont {C.~J.}\ \bibnamefont {Pollock}}, \bibinfo {author} {\bibfnamefont {J.~C.}\ \bibnamefont {Dorelli}}, \bibinfo {author} {\bibfnamefont {L.~A.}\ \bibnamefont {Avanov}}, \bibinfo {author} {\bibfnamefont {B.}~\bibnamefont {Lavraud}}, \bibinfo {author} {\bibfnamefont {Y.}~\bibnamefont {Saito}}, \bibinfo {author} {\bibfnamefont {R.}~\bibnamefont {Nakamura}}, \bibinfo {author} {\bibfnamefont {D.}~\bibnamefont {Fischer}}, \bibinfo {author} {\bibfnamefont {W.}~\bibnamefont {Baumjohann}}, \bibinfo {author}
  {\bibfnamefont {F.}~\bibnamefont {Plaschke}}, \bibinfo {author} {\bibfnamefont {Y.}~\bibnamefont {Narita}}, \bibinfo {author} {\bibfnamefont {W.}~\bibnamefont {Magnes}}, \bibinfo {author} {\bibfnamefont {C.~T.}\ \bibnamefont {Russell}}, \bibinfo {author} {\bibfnamefont {R.~J.}\ \bibnamefont {Strangeway}}, \bibinfo {author} {\bibfnamefont {O.~L.}\ \bibnamefont {Contel}}, \bibinfo {author} {\bibfnamefont {Y.}~\bibnamefont {Khotyaintsev}},\ and\ \bibinfo {author} {\bibfnamefont {F.}~\bibnamefont {Valentini}},\ }\href {https://doi.org/10.1017/S0022377820000021} {\bibfield  {journal} {\bibinfo  {journal} {J. Plasma Phys.}\ }\textbf {\bibinfo {volume} {86}},\ \bibinfo {pages} {905860108} (\bibinfo {year} {2020})}\BibitemShut {NoStop}%
\bibitem [{\citenamefont {Zhdankin}(2022)}]{zhdankin_generalized_2022}%
  \BibitemOpen
  \bibfield  {author} {\bibinfo {author} {\bibfnamefont {V.}~\bibnamefont {Zhdankin}},\ }\href {https://doi.org/10.1103/PhysRevX.12.031011} {\bibfield  {journal} {\bibinfo  {journal} {Phys. Rev. X}\ }\textbf {\bibinfo {volume} {12}},\ \bibinfo {pages} {031011} (\bibinfo {year} {2022})}\BibitemShut {NoStop}%
\bibitem [{\citenamefont {{Sorriso-Valvo}}\ \emph {et~al.}(2019)\citenamefont {{Sorriso-Valvo}}, \citenamefont {Catapano}, \citenamefont {Retin{\`o}}, \citenamefont {Le~Contel}, \citenamefont {Perrone}, \citenamefont {Roberts}, \citenamefont {Coburn}, \citenamefont {Panebianco}, \citenamefont {Valentini}, \citenamefont {Perri}, \citenamefont {Greco}, \citenamefont {Malara}, \citenamefont {Carbone}, \citenamefont {Veltri}, \citenamefont {Pezzi}, \citenamefont {Fraternale}, \citenamefont {Di~Mare}, \citenamefont {Marino}, \citenamefont {Giles}, \citenamefont {Moore}, \citenamefont {Russell}, \citenamefont {Torbert}, \citenamefont {Burch},\ and\ \citenamefont {Khotyaintsev}}]{sorriso-valvo_turbulencedriven_2019}%
  \BibitemOpen
  \bibfield  {author} {\bibinfo {author} {\bibfnamefont {L.}~\bibnamefont {{Sorriso-Valvo}}}, \bibinfo {author} {\bibfnamefont {F.}~\bibnamefont {Catapano}}, \bibinfo {author} {\bibfnamefont {A.}~\bibnamefont {Retin{\`o}}}, \bibinfo {author} {\bibfnamefont {O.}~\bibnamefont {Le~Contel}}, \bibinfo {author} {\bibfnamefont {D.}~\bibnamefont {Perrone}}, \bibinfo {author} {\bibfnamefont {O.~W.}\ \bibnamefont {Roberts}}, \bibinfo {author} {\bibfnamefont {J.~T.}\ \bibnamefont {Coburn}}, \bibinfo {author} {\bibfnamefont {V.}~\bibnamefont {Panebianco}}, \bibinfo {author} {\bibfnamefont {F.}~\bibnamefont {Valentini}}, \bibinfo {author} {\bibfnamefont {S.}~\bibnamefont {Perri}}, \bibinfo {author} {\bibfnamefont {A.}~\bibnamefont {Greco}}, \bibinfo {author} {\bibfnamefont {F.}~\bibnamefont {Malara}}, \bibinfo {author} {\bibfnamefont {V.}~\bibnamefont {Carbone}}, \bibinfo {author} {\bibfnamefont {P.}~\bibnamefont {Veltri}}, \bibinfo {author} {\bibfnamefont {O.}~\bibnamefont {Pezzi}}, \bibinfo {author} {\bibfnamefont
  {F.}~\bibnamefont {Fraternale}}, \bibinfo {author} {\bibfnamefont {F.}~\bibnamefont {Di~Mare}}, \bibinfo {author} {\bibfnamefont {R.}~\bibnamefont {Marino}}, \bibinfo {author} {\bibfnamefont {B.}~\bibnamefont {Giles}}, \bibinfo {author} {\bibfnamefont {T.~E.}\ \bibnamefont {Moore}}, \bibinfo {author} {\bibfnamefont {C.~T.}\ \bibnamefont {Russell}}, \bibinfo {author} {\bibfnamefont {R.~B.}\ \bibnamefont {Torbert}}, \bibinfo {author} {\bibfnamefont {J.~L.}\ \bibnamefont {Burch}},\ and\ \bibinfo {author} {\bibfnamefont {Y.~V.}\ \bibnamefont {Khotyaintsev}},\ }\href {https://doi.org/10.1103/PhysRevLett.122.035102} {\bibfield  {journal} {\bibinfo  {journal} {Phys. Rev. Lett.}\ }\textbf {\bibinfo {volume} {122}},\ \bibinfo {pages} {035102} (\bibinfo {year} {2019})}\BibitemShut {NoStop}%
\bibitem [{\citenamefont {Parks}\ \emph {et~al.}(2012)\citenamefont {Parks}, \citenamefont {Lee}, \citenamefont {McCarthy}, \citenamefont {Goldstein}, \citenamefont {Fu}, \citenamefont {Cao}, \citenamefont {Canu}, \citenamefont {Lin}, \citenamefont {Wilber}, \citenamefont {Dandouras}, \citenamefont {R{\'e}me},\ and\ \citenamefont {Fazakerley}}]{parks_entropy_2012}%
  \BibitemOpen
  \bibfield  {author} {\bibinfo {author} {\bibfnamefont {G.~K.}\ \bibnamefont {Parks}}, \bibinfo {author} {\bibfnamefont {E.}~\bibnamefont {Lee}}, \bibinfo {author} {\bibfnamefont {M.}~\bibnamefont {McCarthy}}, \bibinfo {author} {\bibfnamefont {M.}~\bibnamefont {Goldstein}}, \bibinfo {author} {\bibfnamefont {S.~Y.}\ \bibnamefont {Fu}}, \bibinfo {author} {\bibfnamefont {J.~B.}\ \bibnamefont {Cao}}, \bibinfo {author} {\bibfnamefont {P.}~\bibnamefont {Canu}}, \bibinfo {author} {\bibfnamefont {N.}~\bibnamefont {Lin}}, \bibinfo {author} {\bibfnamefont {M.}~\bibnamefont {Wilber}}, \bibinfo {author} {\bibfnamefont {I.}~\bibnamefont {Dandouras}}, \bibinfo {author} {\bibfnamefont {H.}~\bibnamefont {R{\'e}me}},\ and\ \bibinfo {author} {\bibfnamefont {A.}~\bibnamefont {Fazakerley}},\ }\href {https://doi.org/10.1103/PhysRevLett.108.061102} {\bibfield  {journal} {\bibinfo  {journal} {Phys. Rev. Lett.}\ }\textbf {\bibinfo {volume} {108}},\ \bibinfo {pages} {061102} (\bibinfo {year} {2012})}\BibitemShut {NoStop}%
\bibitem [{\citenamefont {Agapitov}\ \emph {et~al.}(2023)\citenamefont {Agapitov}, \citenamefont {Krasnoselskikh}, \citenamefont {Balikhin}, \citenamefont {Bonnell}, \citenamefont {Mozer},\ and\ \citenamefont {Avanov}}]{agapitov_energy_2023}%
  \BibitemOpen
  \bibfield  {author} {\bibinfo {author} {\bibfnamefont {O.~V.}\ \bibnamefont {Agapitov}}, \bibinfo {author} {\bibfnamefont {V.}~\bibnamefont {Krasnoselskikh}}, \bibinfo {author} {\bibfnamefont {M.}~\bibnamefont {Balikhin}}, \bibinfo {author} {\bibfnamefont {J.~W.}\ \bibnamefont {Bonnell}}, \bibinfo {author} {\bibfnamefont {F.~S.}\ \bibnamefont {Mozer}},\ and\ \bibinfo {author} {\bibfnamefont {L.}~\bibnamefont {Avanov}},\ }\href {https://doi.org/10.3847/1538-4357/acdb7b} {\bibfield  {journal} {\bibinfo  {journal} {Astrophys. J.}\ }\textbf {\bibinfo {volume} {952}},\ \bibinfo {pages} {154} (\bibinfo {year} {2023})}\BibitemShut {NoStop}%
\bibitem [{\citenamefont {Bale}\ \emph {et~al.}(2009)\citenamefont {Bale}, \citenamefont {Kasper}, \citenamefont {Howes}, \citenamefont {Quataert}, \citenamefont {Salem},\ and\ \citenamefont {Sundkvist}}]{bale_magnetic_2009}%
  \BibitemOpen
  \bibfield  {author} {\bibinfo {author} {\bibfnamefont {S.~D.}\ \bibnamefont {Bale}}, \bibinfo {author} {\bibfnamefont {J.~C.}\ \bibnamefont {Kasper}}, \bibinfo {author} {\bibfnamefont {G.~G.}\ \bibnamefont {Howes}}, \bibinfo {author} {\bibfnamefont {E.}~\bibnamefont {Quataert}}, \bibinfo {author} {\bibfnamefont {C.}~\bibnamefont {Salem}},\ and\ \bibinfo {author} {\bibfnamefont {D.}~\bibnamefont {Sundkvist}},\ }\href {https://doi.org/10.1103/PhysRevLett.103.211101} {\bibfield  {journal} {\bibinfo  {journal} {Phys. Rev. Lett.}\ }\textbf {\bibinfo {volume} {103}},\ \bibinfo {pages} {211101} (\bibinfo {year} {2009})}\BibitemShut {NoStop}%
\bibitem [{\citenamefont {Maruca}\ \emph {et~al.}(2018)\citenamefont {Maruca}, \citenamefont {Chasapis}, \citenamefont {Gary}, \citenamefont {Bandyopadhyay}, \citenamefont {Chhiber}, \citenamefont {Parashar}, \citenamefont {Matthaeus}, \citenamefont {Shay}, \citenamefont {Burch}, \citenamefont {Moore}, \citenamefont {Pollock}, \citenamefont {Giles}, \citenamefont {Paterson}, \citenamefont {Dorelli}, \citenamefont {Gershman}, \citenamefont {Torbert}, \citenamefont {Russell},\ and\ \citenamefont {Strangeway}}]{maruca_mms_2018}%
  \BibitemOpen
  \bibfield  {author} {\bibinfo {author} {\bibfnamefont {B.~A.}\ \bibnamefont {Maruca}}, \bibinfo {author} {\bibfnamefont {A.}~\bibnamefont {Chasapis}}, \bibinfo {author} {\bibfnamefont {S.~P.}\ \bibnamefont {Gary}}, \bibinfo {author} {\bibfnamefont {R.}~\bibnamefont {Bandyopadhyay}}, \bibinfo {author} {\bibfnamefont {R.}~\bibnamefont {Chhiber}}, \bibinfo {author} {\bibfnamefont {T.~N.}\ \bibnamefont {Parashar}}, \bibinfo {author} {\bibfnamefont {W.~H.}\ \bibnamefont {Matthaeus}}, \bibinfo {author} {\bibfnamefont {M.~A.}\ \bibnamefont {Shay}}, \bibinfo {author} {\bibfnamefont {J.~L.}\ \bibnamefont {Burch}}, \bibinfo {author} {\bibfnamefont {T.~E.}\ \bibnamefont {Moore}}, \bibinfo {author} {\bibfnamefont {C.~J.}\ \bibnamefont {Pollock}}, \bibinfo {author} {\bibfnamefont {B.~J.}\ \bibnamefont {Giles}}, \bibinfo {author} {\bibfnamefont {W.~R.}\ \bibnamefont {Paterson}}, \bibinfo {author} {\bibfnamefont {J.}~\bibnamefont {Dorelli}}, \bibinfo {author} {\bibfnamefont {D.~J.}\ \bibnamefont {Gershman}}, \bibinfo
  {author} {\bibfnamefont {R.~B.}\ \bibnamefont {Torbert}}, \bibinfo {author} {\bibfnamefont {C.~T.}\ \bibnamefont {Russell}},\ and\ \bibinfo {author} {\bibfnamefont {R.~J.}\ \bibnamefont {Strangeway}},\ }\href {https://doi.org/10.3847/1538-4357/aaddfb} {\bibfield  {journal} {\bibinfo  {journal} {Astrophys. J.}\ }\textbf {\bibinfo {volume} {866}},\ \bibinfo {pages} {25} (\bibinfo {year} {2018})}\BibitemShut {NoStop}%
\bibitem [{\citenamefont {Martinovi{\'c}}\ \emph {et~al.}(2020)\citenamefont {Martinovi{\'c}}, \citenamefont {Klein}, \citenamefont {Kasper}, \citenamefont {Case}, \citenamefont {Korreck}, \citenamefont {Larson}, \citenamefont {Livi}, \citenamefont {Stevens}, \citenamefont {Whittlesey}, \citenamefont {Chandran}, \citenamefont {Alterman}, \citenamefont {Huang}, \citenamefont {Chen}, \citenamefont {Bale}, \citenamefont {Pulupa}, \citenamefont {Malaspina}, \citenamefont {Bonnell}, \citenamefont {Harvey}, \citenamefont {Goetz}, \citenamefont {Dudok De~Wit},\ and\ \citenamefont {MacDowall}}]{martinovic_enhancement_2020}%
  \BibitemOpen
  \bibfield  {author} {\bibinfo {author} {\bibfnamefont {M.~M.}\ \bibnamefont {Martinovi{\'c}}}, \bibinfo {author} {\bibfnamefont {K.~G.}\ \bibnamefont {Klein}}, \bibinfo {author} {\bibfnamefont {J.~C.}\ \bibnamefont {Kasper}}, \bibinfo {author} {\bibfnamefont {A.~W.}\ \bibnamefont {Case}}, \bibinfo {author} {\bibfnamefont {K.~E.}\ \bibnamefont {Korreck}}, \bibinfo {author} {\bibfnamefont {D.}~\bibnamefont {Larson}}, \bibinfo {author} {\bibfnamefont {R.}~\bibnamefont {Livi}}, \bibinfo {author} {\bibfnamefont {M.}~\bibnamefont {Stevens}}, \bibinfo {author} {\bibfnamefont {P.}~\bibnamefont {Whittlesey}}, \bibinfo {author} {\bibfnamefont {B.~D.~G.}\ \bibnamefont {Chandran}}, \bibinfo {author} {\bibfnamefont {B.~L.}\ \bibnamefont {Alterman}}, \bibinfo {author} {\bibfnamefont {J.}~\bibnamefont {Huang}}, \bibinfo {author} {\bibfnamefont {C.~H.~K.}\ \bibnamefont {Chen}}, \bibinfo {author} {\bibfnamefont {S.~D.}\ \bibnamefont {Bale}}, \bibinfo {author} {\bibfnamefont {M.}~\bibnamefont {Pulupa}}, \bibinfo {author}
  {\bibfnamefont {D.~M.}\ \bibnamefont {Malaspina}}, \bibinfo {author} {\bibfnamefont {J.~W.}\ \bibnamefont {Bonnell}}, \bibinfo {author} {\bibfnamefont {P.~R.}\ \bibnamefont {Harvey}}, \bibinfo {author} {\bibfnamefont {K.}~\bibnamefont {Goetz}}, \bibinfo {author} {\bibfnamefont {T.}~\bibnamefont {Dudok De~Wit}},\ and\ \bibinfo {author} {\bibfnamefont {R.~J.}\ \bibnamefont {MacDowall}},\ }\href {https://doi.org/10.3847/1538-4365/ab527f} {\bibfield  {journal} {\bibinfo  {journal} {ApJS}\ }\textbf {\bibinfo {volume} {246}},\ \bibinfo {pages} {30} (\bibinfo {year} {2020})}\BibitemShut {NoStop}%
\bibitem [{\citenamefont {Verniero}\ \emph {et~al.}(2022)\citenamefont {Verniero}, \citenamefont {Chandran}, \citenamefont {Larson}, \citenamefont {Paulson}, \citenamefont {Alterman}, \citenamefont {Badman}, \citenamefont {Bale}, \citenamefont {Bonnell}, \citenamefont {Bowen}, \citenamefont {De~Wit}, \citenamefont {Kasper}, \citenamefont {Klein}, \citenamefont {Lichko}, \citenamefont {Livi}, \citenamefont {McManus}, \citenamefont {Rahmati}, \citenamefont {Verscharen}, \citenamefont {Walters},\ and\ \citenamefont {Whittlesey}}]{verniero_strong_2022}%
  \BibitemOpen
  \bibfield  {author} {\bibinfo {author} {\bibfnamefont {J.~L.}\ \bibnamefont {Verniero}}, \bibinfo {author} {\bibfnamefont {B.~D.~G.}\ \bibnamefont {Chandran}}, \bibinfo {author} {\bibfnamefont {D.~E.}\ \bibnamefont {Larson}}, \bibinfo {author} {\bibfnamefont {K.}~\bibnamefont {Paulson}}, \bibinfo {author} {\bibfnamefont {B.~L.}\ \bibnamefont {Alterman}}, \bibinfo {author} {\bibfnamefont {S.}~\bibnamefont {Badman}}, \bibinfo {author} {\bibfnamefont {S.~D.}\ \bibnamefont {Bale}}, \bibinfo {author} {\bibfnamefont {J.~W.}\ \bibnamefont {Bonnell}}, \bibinfo {author} {\bibfnamefont {T.~A.}\ \bibnamefont {Bowen}}, \bibinfo {author} {\bibfnamefont {T.~D.}\ \bibnamefont {De~Wit}}, \bibinfo {author} {\bibfnamefont {J.~C.}\ \bibnamefont {Kasper}}, \bibinfo {author} {\bibfnamefont {K.~G.}\ \bibnamefont {Klein}}, \bibinfo {author} {\bibfnamefont {E.}~\bibnamefont {Lichko}}, \bibinfo {author} {\bibfnamefont {R.}~\bibnamefont {Livi}}, \bibinfo {author} {\bibfnamefont {M.~D.}\ \bibnamefont {McManus}}, \bibinfo {author}
  {\bibfnamefont {A.}~\bibnamefont {Rahmati}}, \bibinfo {author} {\bibfnamefont {D.}~\bibnamefont {Verscharen}}, \bibinfo {author} {\bibfnamefont {J.}~\bibnamefont {Walters}},\ and\ \bibinfo {author} {\bibfnamefont {P.~L.}\ \bibnamefont {Whittlesey}},\ }\href {https://doi.org/10.3847/1538-4357/ac36d5} {\bibfield  {journal} {\bibinfo  {journal} {ApJ}\ }\textbf {\bibinfo {volume} {924}},\ \bibinfo {pages} {112} (\bibinfo {year} {2022})}\BibitemShut {NoStop}%
\bibitem [{\citenamefont {Ewart}\ \emph {et~al.}(2022)\citenamefont {Ewart}, \citenamefont {Brown}, \citenamefont {Adkins},\ and\ \citenamefont {Schekochihin}}]{ewart_collisionless_2022}%
  \BibitemOpen
  \bibfield  {author} {\bibinfo {author} {\bibfnamefont {R.}~\bibnamefont {Ewart}}, \bibinfo {author} {\bibfnamefont {A.}~\bibnamefont {Brown}}, \bibinfo {author} {\bibfnamefont {T.}~\bibnamefont {Adkins}},\ and\ \bibinfo {author} {\bibfnamefont {A.}~\bibnamefont {Schekochihin}},\ }\href {https://doi.org/10.1017/S0022377822000782} {\bibfield  {journal} {\bibinfo  {journal} {J. Plasma Phys.}\ }\textbf {\bibinfo {volume} {88}},\ \bibinfo {pages} {925880501} (\bibinfo {year} {2022})}\BibitemShut {NoStop}%
\bibitem [{\citenamefont {Ewart}\ \emph {et~al.}(2023)\citenamefont {Ewart}, \citenamefont {Nastac},\ and\ \citenamefont {Schekochihin}}]{ewart_nonthermal_2023}%
  \BibitemOpen
  \bibfield  {author} {\bibinfo {author} {\bibfnamefont {R.}~\bibnamefont {Ewart}}, \bibinfo {author} {\bibfnamefont {M.}~\bibnamefont {Nastac}},\ and\ \bibinfo {author} {\bibfnamefont {A.}~\bibnamefont {Schekochihin}},\ }\href {https://doi.org/10.1017/S0022377823000983} {\bibfield  {journal} {\bibinfo  {journal} {J. Plasma Phys.}\ }\textbf {\bibinfo {volume} {89}},\ \bibinfo {pages} {905890516} (\bibinfo {year} {2023})}\BibitemShut {NoStop}%
\bibitem [{\citenamefont {Ewart}\ \emph {et~al.}(2025)\citenamefont {Ewart}, \citenamefont {Nastac}, \citenamefont {Bilbao}, \citenamefont {Silva}, \citenamefont {Silva},\ and\ \citenamefont {Schekochihin}}]{ewart_relaxation_2025}%
  \BibitemOpen
  \bibfield  {author} {\bibinfo {author} {\bibfnamefont {R.~J.}\ \bibnamefont {Ewart}}, \bibinfo {author} {\bibfnamefont {M.~L.}\ \bibnamefont {Nastac}}, \bibinfo {author} {\bibfnamefont {P.~J.}\ \bibnamefont {Bilbao}}, \bibinfo {author} {\bibfnamefont {T.}~\bibnamefont {Silva}}, \bibinfo {author} {\bibfnamefont {L.~O.}\ \bibnamefont {Silva}},\ and\ \bibinfo {author} {\bibfnamefont {A.~A.}\ \bibnamefont {Schekochihin}},\ }\href {https://doi.org/10.1073/pnas.2417813122} {\bibfield  {journal} {\bibinfo  {journal} {Proc. Natl. Acad. Sci. U.S.A.}\ }\textbf {\bibinfo {volume} {122}},\ \bibinfo {pages} {e2417813122} (\bibinfo {year} {2025})}\BibitemShut {NoStop}%
\bibitem [{\citenamefont {Walters}\ \emph {et~al.}(2023)\citenamefont {Walters}, \citenamefont {Klein}, \citenamefont {Lichko}, \citenamefont {Stevens}, \citenamefont {Verscharen},\ and\ \citenamefont {Chandran}}]{walters_effects_2023}%
  \BibitemOpen
  \bibfield  {author} {\bibinfo {author} {\bibfnamefont {J.}~\bibnamefont {Walters}}, \bibinfo {author} {\bibfnamefont {K.~G.}\ \bibnamefont {Klein}}, \bibinfo {author} {\bibfnamefont {E.}~\bibnamefont {Lichko}}, \bibinfo {author} {\bibfnamefont {M.~L.}\ \bibnamefont {Stevens}}, \bibinfo {author} {\bibfnamefont {D.}~\bibnamefont {Verscharen}},\ and\ \bibinfo {author} {\bibfnamefont {B.~D.~G.}\ \bibnamefont {Chandran}},\ }\href {https://doi.org/10.3847/1538-4357/acf1fa} {\bibfield  {journal} {\bibinfo  {journal} {ApJ}\ }\textbf {\bibinfo {volume} {955}},\ \bibinfo {pages} {97} (\bibinfo {year} {2023})}\BibitemShut {NoStop}%
\bibitem [{\citenamefont {Stix}(1992)}]{stix_waves_1992}%
  \BibitemOpen
  \bibfield  {author} {\bibinfo {author} {\bibfnamefont {T.~H.}\ \bibnamefont {Stix}},\ }\href@noop {} {\emph {\bibinfo {title} {Waves in {{Plasmas}}}}},\ \bibinfo {edition} {1st}\ ed.\ (\bibinfo  {publisher} {American Institute of Physics},\ \bibinfo {address} {Melville, NY},\ \bibinfo {year} {1992})\BibitemShut {NoStop}%
\bibitem [{\citenamefont {Maron}\ and\ \citenamefont {Goldreich}(2001)}]{maron_simulations_2001}%
  \BibitemOpen
  \bibfield  {author} {\bibinfo {author} {\bibfnamefont {J.}~\bibnamefont {Maron}}\ and\ \bibinfo {author} {\bibfnamefont {P.}~\bibnamefont {Goldreich}},\ }\href {https://doi.org/10.1086/321413} {\bibfield  {journal} {\bibinfo  {journal} {The Astrophysical Journal}\ }\textbf {\bibinfo {volume} {554}},\ \bibinfo {pages} {1175} (\bibinfo {year} {2001})}\BibitemShut {NoStop}%
\bibitem [{\citenamefont {Matthaeus}\ \emph {et~al.}(2014)\citenamefont {Matthaeus}, \citenamefont {Oughton}, \citenamefont {Osman}, \citenamefont {Servidio}, \citenamefont {Wan}, \citenamefont {Gary}, \citenamefont {Shay}, \citenamefont {Valentini}, \citenamefont {Roytershteyn}, \citenamefont {Karimabadi},\ and\ \citenamefont {Chapman}}]{matthaeus_nonlinear_2014}%
  \BibitemOpen
  \bibfield  {author} {\bibinfo {author} {\bibfnamefont {W.~H.}\ \bibnamefont {Matthaeus}}, \bibinfo {author} {\bibfnamefont {S.}~\bibnamefont {Oughton}}, \bibinfo {author} {\bibfnamefont {K.~T.}\ \bibnamefont {Osman}}, \bibinfo {author} {\bibfnamefont {S.}~\bibnamefont {Servidio}}, \bibinfo {author} {\bibfnamefont {M.}~\bibnamefont {Wan}}, \bibinfo {author} {\bibfnamefont {S.~P.}\ \bibnamefont {Gary}}, \bibinfo {author} {\bibfnamefont {M.~A.}\ \bibnamefont {Shay}}, \bibinfo {author} {\bibfnamefont {F.}~\bibnamefont {Valentini}}, \bibinfo {author} {\bibfnamefont {V.}~\bibnamefont {Roytershteyn}}, \bibinfo {author} {\bibfnamefont {H.}~\bibnamefont {Karimabadi}},\ and\ \bibinfo {author} {\bibfnamefont {S.~C.}\ \bibnamefont {Chapman}},\ }\href {https://doi.org/10.1088/0004-637X/790/2/155} {\bibfield  {journal} {\bibinfo  {journal} {Astrophys. J.}\ }\textbf {\bibinfo {volume} {790}},\ \bibinfo {pages} {155} (\bibinfo {year} {2014})}\BibitemShut {NoStop}%
\bibitem [{\citenamefont {Li}\ \emph {et~al.}(2022)\citenamefont {Li}, \citenamefont {Liu}, \citenamefont {Zhou}, \citenamefont {Li}, \citenamefont {Omura}, \citenamefont {Yue}, \citenamefont {Zong}, \citenamefont {Pu}, \citenamefont {Fu}, \citenamefont {Xie}, \citenamefont {Russell}, \citenamefont {Pollock}, \citenamefont {Le},\ and\ \citenamefont {Burch}}]{li_anomalous_2022}%
  \BibitemOpen
  \bibfield  {author} {\bibinfo {author} {\bibfnamefont {J.-H.}\ \bibnamefont {Li}}, \bibinfo {author} {\bibfnamefont {Z.-Y.}\ \bibnamefont {Liu}}, \bibinfo {author} {\bibfnamefont {X.-Z.}\ \bibnamefont {Zhou}}, \bibinfo {author} {\bibfnamefont {L.}~\bibnamefont {Li}}, \bibinfo {author} {\bibfnamefont {Y.}~\bibnamefont {Omura}}, \bibinfo {author} {\bibfnamefont {C.}~\bibnamefont {Yue}}, \bibinfo {author} {\bibfnamefont {Q.-G.}\ \bibnamefont {Zong}}, \bibinfo {author} {\bibfnamefont {Z.-Y.}\ \bibnamefont {Pu}}, \bibinfo {author} {\bibfnamefont {S.-Y.}\ \bibnamefont {Fu}}, \bibinfo {author} {\bibfnamefont {L.}~\bibnamefont {Xie}}, \bibinfo {author} {\bibfnamefont {C.~T.}\ \bibnamefont {Russell}}, \bibinfo {author} {\bibfnamefont {C.~J.}\ \bibnamefont {Pollock}}, \bibinfo {author} {\bibfnamefont {G.}~\bibnamefont {Le}},\ and\ \bibinfo {author} {\bibfnamefont {J.~L.}\ \bibnamefont {Burch}},\ }\bibfield  {journal} {\bibinfo  {journal} {Commun Phys}\ }\textbf {\bibinfo {volume} {5}},\ \href
  {https://doi.org/10.1038/s42005-022-01083-y} {10.1038/s42005-022-01083-y} (\bibinfo {year} {2022})\BibitemShut {NoStop}%
\bibitem [{\citenamefont {Chaston}\ \emph {et~al.}(2020)\citenamefont {Chaston}, \citenamefont {Travnicek},\ and\ \citenamefont {Russell}}]{chaston_turbulent_2020}%
  \BibitemOpen
  \bibfield  {author} {\bibinfo {author} {\bibfnamefont {C.~C.}\ \bibnamefont {Chaston}}, \bibinfo {author} {\bibfnamefont {P.}~\bibnamefont {Travnicek}},\ and\ \bibinfo {author} {\bibfnamefont {C.~T.}\ \bibnamefont {Russell}},\ }\href {https://doi.org/10.1029/2020GL089613} {\bibfield  {journal} {\bibinfo  {journal} {Geophys. Res. Lett.}\ }\textbf {\bibinfo {volume} {47}},\ \bibinfo {pages} {e2020GL089613} (\bibinfo {year} {2020})}\BibitemShut {NoStop}%
\bibitem [{\citenamefont {Chaston}\ and\ \citenamefont {Travnicek}(2021)}]{chaston_ion_2021}%
  \BibitemOpen
  \bibfield  {author} {\bibinfo {author} {\bibfnamefont {C.~C.}\ \bibnamefont {Chaston}}\ and\ \bibinfo {author} {\bibfnamefont {P.}~\bibnamefont {Travnicek}},\ }\href {https://doi.org/10.1029/2021GL094029} {\bibfield  {journal} {\bibinfo  {journal} {Geophys. Res. Lett.}\ }\textbf {\bibinfo {volume} {48}},\ \bibinfo {pages} {e2021GL094029} (\bibinfo {year} {2021})}\BibitemShut {NoStop}%
\bibitem [{\citenamefont {Artemyev}\ \emph {et~al.}(2020)\citenamefont {Artemyev}, \citenamefont {Neishtadt}, \citenamefont {Vasiliev}, \citenamefont {Angelopoulos}, \citenamefont {Vinogradov},\ and\ \citenamefont {Zelenyi}}]{artemyev_superfast_2020}%
  \BibitemOpen
  \bibfield  {author} {\bibinfo {author} {\bibfnamefont {A.~V.}\ \bibnamefont {Artemyev}}, \bibinfo {author} {\bibfnamefont {A.~I.}\ \bibnamefont {Neishtadt}}, \bibinfo {author} {\bibfnamefont {A.~A.}\ \bibnamefont {Vasiliev}}, \bibinfo {author} {\bibfnamefont {V.}~\bibnamefont {Angelopoulos}}, \bibinfo {author} {\bibfnamefont {A.~A.}\ \bibnamefont {Vinogradov}},\ and\ \bibinfo {author} {\bibfnamefont {L.~M.}\ \bibnamefont {Zelenyi}},\ }\href {https://doi.org/10.1103/PhysRevE.102.033201} {\bibfield  {journal} {\bibinfo  {journal} {Phys. Rev. E}\ }\textbf {\bibinfo {volume} {102}},\ \bibinfo {pages} {033201} (\bibinfo {year} {2020})}\BibitemShut {NoStop}%
\bibitem [{\citenamefont {Lemoine}(2023)}]{lemoine_particle_2023}%
  \BibitemOpen
  \bibfield  {author} {\bibinfo {author} {\bibfnamefont {M.}~\bibnamefont {Lemoine}},\ }\href {https://doi.org/10.1017/S0022377823000946} {\bibfield  {journal} {\bibinfo  {journal} {J. Plasma Phys.}\ }\textbf {\bibinfo {volume} {89}},\ \bibinfo {pages} {175890501} (\bibinfo {year} {2023})}\BibitemShut {NoStop}%
\bibitem [{\citenamefont {Russell}\ \emph {et~al.}(2016)\citenamefont {Russell}, \citenamefont {Anderson}, \citenamefont {Baumjohann}, \citenamefont {Bromund}, \citenamefont {Dearborn}, \citenamefont {Fischer}, \citenamefont {Le}, \citenamefont {Leinweber}, \citenamefont {Leneman}, \citenamefont {Magnes}, \citenamefont {Means}, \citenamefont {Moldwin}, \citenamefont {Nakamura}, \citenamefont {Pierce}, \citenamefont {Plaschke}, \citenamefont {Rowe}, \citenamefont {Slavin}, \citenamefont {Strangeway}, \citenamefont {Torbert}, \citenamefont {Hagen}, \citenamefont {Jernej}, \citenamefont {Valavanoglou},\ and\ \citenamefont {Richter}}]{russell_magnetospheric_2016}%
  \BibitemOpen
  \bibfield  {author} {\bibinfo {author} {\bibfnamefont {C.~T.}\ \bibnamefont {Russell}}, \bibinfo {author} {\bibfnamefont {B.~J.}\ \bibnamefont {Anderson}}, \bibinfo {author} {\bibfnamefont {W.}~\bibnamefont {Baumjohann}}, \bibinfo {author} {\bibfnamefont {K.~R.}\ \bibnamefont {Bromund}}, \bibinfo {author} {\bibfnamefont {D.}~\bibnamefont {Dearborn}}, \bibinfo {author} {\bibfnamefont {D.}~\bibnamefont {Fischer}}, \bibinfo {author} {\bibfnamefont {G.}~\bibnamefont {Le}}, \bibinfo {author} {\bibfnamefont {H.~K.}\ \bibnamefont {Leinweber}}, \bibinfo {author} {\bibfnamefont {D.}~\bibnamefont {Leneman}}, \bibinfo {author} {\bibfnamefont {W.}~\bibnamefont {Magnes}}, \bibinfo {author} {\bibfnamefont {J.~D.}\ \bibnamefont {Means}}, \bibinfo {author} {\bibfnamefont {M.~B.}\ \bibnamefont {Moldwin}}, \bibinfo {author} {\bibfnamefont {R.}~\bibnamefont {Nakamura}}, \bibinfo {author} {\bibfnamefont {D.}~\bibnamefont {Pierce}}, \bibinfo {author} {\bibfnamefont {F.}~\bibnamefont {Plaschke}}, \bibinfo {author} {\bibfnamefont
  {K.~M.}\ \bibnamefont {Rowe}}, \bibinfo {author} {\bibfnamefont {J.~A.}\ \bibnamefont {Slavin}}, \bibinfo {author} {\bibfnamefont {R.~J.}\ \bibnamefont {Strangeway}}, \bibinfo {author} {\bibfnamefont {R.}~\bibnamefont {Torbert}}, \bibinfo {author} {\bibfnamefont {C.}~\bibnamefont {Hagen}}, \bibinfo {author} {\bibfnamefont {I.}~\bibnamefont {Jernej}}, \bibinfo {author} {\bibfnamefont {A.}~\bibnamefont {Valavanoglou}},\ and\ \bibinfo {author} {\bibfnamefont {I.}~\bibnamefont {Richter}},\ }\href {https://doi.org/10.1007/s11214-014-0057-3} {\bibfield  {journal} {\bibinfo  {journal} {Space Sci. Rev.}\ }\textbf {\bibinfo {volume} {199}},\ \bibinfo {pages} {189} (\bibinfo {year} {2016})}\BibitemShut {NoStop}%
\bibitem [{\citenamefont {Pollock}\ \emph {et~al.}(2016)\citenamefont {Pollock}, \citenamefont {Moore}, \citenamefont {Jacques}, \citenamefont {Burch}, \citenamefont {Gliese}, \citenamefont {Saito}, \citenamefont {Omoto}, \citenamefont {Avanov}, \citenamefont {Barrie}, \citenamefont {Coffey}, \citenamefont {Dorelli}, \citenamefont {Gershman}, \citenamefont {Giles}, \citenamefont {Rosnack}, \citenamefont {Salo}, \citenamefont {Yokota}, \citenamefont {Adrian}, \citenamefont {Aoustin}, \citenamefont {Auletti}, \citenamefont {Aung}, \citenamefont {Bigio}, \citenamefont {Cao}, \citenamefont {Chandler}, \citenamefont {Chornay}, \citenamefont {Christian}, \citenamefont {Clark}, \citenamefont {Collinson}, \citenamefont {Corris}, \citenamefont {De~Los~Santos}, \citenamefont {Devlin}, \citenamefont {Diaz}, \citenamefont {Dickerson}, \citenamefont {Dickson}, \citenamefont {Diekmann}, \citenamefont {Diggs}, \citenamefont {Duncan}, \citenamefont {{Figueroa-Vinas}}, \citenamefont {Firman}, \citenamefont {Freeman},
  \citenamefont {Galassi}, \citenamefont {Garcia}, \citenamefont {Goodhart}, \citenamefont {Guererro}, \citenamefont {Hageman}, \citenamefont {Hanley}, \citenamefont {Hemminger}, \citenamefont {Holland}, \citenamefont {Hutchins}, \citenamefont {James}, \citenamefont {Jones}, \citenamefont {Kreisler}, \citenamefont {Kujawski}, \citenamefont {Lavu}, \citenamefont {Lobell}, \citenamefont {LeCompte}, \citenamefont {Lukemire}, \citenamefont {MacDonald}, \citenamefont {Mariano}, \citenamefont {Mukai}, \citenamefont {Narayanan}, \citenamefont {Nguyan}, \citenamefont {Onizuka}, \citenamefont {Paterson}, \citenamefont {Persyn}, \citenamefont {Piepgrass}, \citenamefont {Cheney}, \citenamefont {Rager}, \citenamefont {Raghuram}, \citenamefont {Ramil}, \citenamefont {Reichenthal}, \citenamefont {Rodriguez}, \citenamefont {Rouzaud}, \citenamefont {Rucker}, \citenamefont {Saito}, \citenamefont {Samara}, \citenamefont {Sauvaud}, \citenamefont {Schuster}, \citenamefont {Shappirio}, \citenamefont {Shelton}, \citenamefont
  {Sher}, \citenamefont {Smith}, \citenamefont {Smith}, \citenamefont {Smith}, \citenamefont {Steinfeld}, \citenamefont {Szymkiewicz}, \citenamefont {Tanimoto}, \citenamefont {Taylor}, \citenamefont {Tucker}, \citenamefont {Tull}, \citenamefont {Uhl}, \citenamefont {Vloet}, \citenamefont {Walpole}, \citenamefont {Weidner}, \citenamefont {White}, \citenamefont {Winkert}, \citenamefont {Yeh},\ and\ \citenamefont {Zeuch}}]{pollock_fast_2016}%
  \BibitemOpen
  \bibfield  {author} {\bibinfo {author} {\bibfnamefont {C.}~\bibnamefont {Pollock}}, \bibinfo {author} {\bibfnamefont {T.}~\bibnamefont {Moore}}, \bibinfo {author} {\bibfnamefont {A.}~\bibnamefont {Jacques}}, \bibinfo {author} {\bibfnamefont {J.}~\bibnamefont {Burch}}, \bibinfo {author} {\bibfnamefont {U.}~\bibnamefont {Gliese}}, \bibinfo {author} {\bibfnamefont {Y.}~\bibnamefont {Saito}}, \bibinfo {author} {\bibfnamefont {T.}~\bibnamefont {Omoto}}, \bibinfo {author} {\bibfnamefont {L.}~\bibnamefont {Avanov}}, \bibinfo {author} {\bibfnamefont {A.}~\bibnamefont {Barrie}}, \bibinfo {author} {\bibfnamefont {V.}~\bibnamefont {Coffey}}, \bibinfo {author} {\bibfnamefont {J.}~\bibnamefont {Dorelli}}, \bibinfo {author} {\bibfnamefont {D.}~\bibnamefont {Gershman}}, \bibinfo {author} {\bibfnamefont {B.}~\bibnamefont {Giles}}, \bibinfo {author} {\bibfnamefont {T.}~\bibnamefont {Rosnack}}, \bibinfo {author} {\bibfnamefont {C.}~\bibnamefont {Salo}}, \bibinfo {author} {\bibfnamefont {S.}~\bibnamefont {Yokota}}, \bibinfo
  {author} {\bibfnamefont {M.}~\bibnamefont {Adrian}}, \bibinfo {author} {\bibfnamefont {C.}~\bibnamefont {Aoustin}}, \bibinfo {author} {\bibfnamefont {C.}~\bibnamefont {Auletti}}, \bibinfo {author} {\bibfnamefont {S.}~\bibnamefont {Aung}}, \bibinfo {author} {\bibfnamefont {V.}~\bibnamefont {Bigio}}, \bibinfo {author} {\bibfnamefont {N.}~\bibnamefont {Cao}}, \bibinfo {author} {\bibfnamefont {M.}~\bibnamefont {Chandler}}, \bibinfo {author} {\bibfnamefont {D.}~\bibnamefont {Chornay}}, \bibinfo {author} {\bibfnamefont {K.}~\bibnamefont {Christian}}, \bibinfo {author} {\bibfnamefont {G.}~\bibnamefont {Clark}}, \bibinfo {author} {\bibfnamefont {G.}~\bibnamefont {Collinson}}, \bibinfo {author} {\bibfnamefont {T.}~\bibnamefont {Corris}}, \bibinfo {author} {\bibfnamefont {A.}~\bibnamefont {De~Los~Santos}}, \bibinfo {author} {\bibfnamefont {R.}~\bibnamefont {Devlin}}, \bibinfo {author} {\bibfnamefont {T.}~\bibnamefont {Diaz}}, \bibinfo {author} {\bibfnamefont {T.}~\bibnamefont {Dickerson}}, \bibinfo {author}
  {\bibfnamefont {C.}~\bibnamefont {Dickson}}, \bibinfo {author} {\bibfnamefont {A.}~\bibnamefont {Diekmann}}, \bibinfo {author} {\bibfnamefont {F.}~\bibnamefont {Diggs}}, \bibinfo {author} {\bibfnamefont {C.}~\bibnamefont {Duncan}}, \bibinfo {author} {\bibfnamefont {A.}~\bibnamefont {{Figueroa-Vinas}}}, \bibinfo {author} {\bibfnamefont {C.}~\bibnamefont {Firman}}, \bibinfo {author} {\bibfnamefont {M.}~\bibnamefont {Freeman}}, \bibinfo {author} {\bibfnamefont {N.}~\bibnamefont {Galassi}}, \bibinfo {author} {\bibfnamefont {K.}~\bibnamefont {Garcia}}, \bibinfo {author} {\bibfnamefont {G.}~\bibnamefont {Goodhart}}, \bibinfo {author} {\bibfnamefont {D.}~\bibnamefont {Guererro}}, \bibinfo {author} {\bibfnamefont {J.}~\bibnamefont {Hageman}}, \bibinfo {author} {\bibfnamefont {J.}~\bibnamefont {Hanley}}, \bibinfo {author} {\bibfnamefont {E.}~\bibnamefont {Hemminger}}, \bibinfo {author} {\bibfnamefont {M.}~\bibnamefont {Holland}}, \bibinfo {author} {\bibfnamefont {M.}~\bibnamefont {Hutchins}}, \bibinfo {author}
  {\bibfnamefont {T.}~\bibnamefont {James}}, \bibinfo {author} {\bibfnamefont {W.}~\bibnamefont {Jones}}, \bibinfo {author} {\bibfnamefont {S.}~\bibnamefont {Kreisler}}, \bibinfo {author} {\bibfnamefont {J.}~\bibnamefont {Kujawski}}, \bibinfo {author} {\bibfnamefont {V.}~\bibnamefont {Lavu}}, \bibinfo {author} {\bibfnamefont {J.}~\bibnamefont {Lobell}}, \bibinfo {author} {\bibfnamefont {E.}~\bibnamefont {LeCompte}}, \bibinfo {author} {\bibfnamefont {A.}~\bibnamefont {Lukemire}}, \bibinfo {author} {\bibfnamefont {E.}~\bibnamefont {MacDonald}}, \bibinfo {author} {\bibfnamefont {A.}~\bibnamefont {Mariano}}, \bibinfo {author} {\bibfnamefont {T.}~\bibnamefont {Mukai}}, \bibinfo {author} {\bibfnamefont {K.}~\bibnamefont {Narayanan}}, \bibinfo {author} {\bibfnamefont {Q.}~\bibnamefont {Nguyan}}, \bibinfo {author} {\bibfnamefont {M.}~\bibnamefont {Onizuka}}, \bibinfo {author} {\bibfnamefont {W.}~\bibnamefont {Paterson}}, \bibinfo {author} {\bibfnamefont {S.}~\bibnamefont {Persyn}}, \bibinfo {author} {\bibfnamefont
  {B.}~\bibnamefont {Piepgrass}}, \bibinfo {author} {\bibfnamefont {F.}~\bibnamefont {Cheney}}, \bibinfo {author} {\bibfnamefont {A.}~\bibnamefont {Rager}}, \bibinfo {author} {\bibfnamefont {T.}~\bibnamefont {Raghuram}}, \bibinfo {author} {\bibfnamefont {A.}~\bibnamefont {Ramil}}, \bibinfo {author} {\bibfnamefont {L.}~\bibnamefont {Reichenthal}}, \bibinfo {author} {\bibfnamefont {H.}~\bibnamefont {Rodriguez}}, \bibinfo {author} {\bibfnamefont {J.}~\bibnamefont {Rouzaud}}, \bibinfo {author} {\bibfnamefont {A.}~\bibnamefont {Rucker}}, \bibinfo {author} {\bibfnamefont {Y.}~\bibnamefont {Saito}}, \bibinfo {author} {\bibfnamefont {M.}~\bibnamefont {Samara}}, \bibinfo {author} {\bibfnamefont {J.-A.}\ \bibnamefont {Sauvaud}}, \bibinfo {author} {\bibfnamefont {D.}~\bibnamefont {Schuster}}, \bibinfo {author} {\bibfnamefont {M.}~\bibnamefont {Shappirio}}, \bibinfo {author} {\bibfnamefont {K.}~\bibnamefont {Shelton}}, \bibinfo {author} {\bibfnamefont {D.}~\bibnamefont {Sher}}, \bibinfo {author} {\bibfnamefont
  {D.}~\bibnamefont {Smith}}, \bibinfo {author} {\bibfnamefont {K.}~\bibnamefont {Smith}}, \bibinfo {author} {\bibfnamefont {S.}~\bibnamefont {Smith}}, \bibinfo {author} {\bibfnamefont {D.}~\bibnamefont {Steinfeld}}, \bibinfo {author} {\bibfnamefont {R.}~\bibnamefont {Szymkiewicz}}, \bibinfo {author} {\bibfnamefont {K.}~\bibnamefont {Tanimoto}}, \bibinfo {author} {\bibfnamefont {J.}~\bibnamefont {Taylor}}, \bibinfo {author} {\bibfnamefont {C.}~\bibnamefont {Tucker}}, \bibinfo {author} {\bibfnamefont {K.}~\bibnamefont {Tull}}, \bibinfo {author} {\bibfnamefont {A.}~\bibnamefont {Uhl}}, \bibinfo {author} {\bibfnamefont {J.}~\bibnamefont {Vloet}}, \bibinfo {author} {\bibfnamefont {P.}~\bibnamefont {Walpole}}, \bibinfo {author} {\bibfnamefont {S.}~\bibnamefont {Weidner}}, \bibinfo {author} {\bibfnamefont {D.}~\bibnamefont {White}}, \bibinfo {author} {\bibfnamefont {G.}~\bibnamefont {Winkert}}, \bibinfo {author} {\bibfnamefont {P.-S.}\ \bibnamefont {Yeh}},\ and\ \bibinfo {author} {\bibfnamefont {M.}~\bibnamefont
  {Zeuch}},\ }\href {https://doi.org/10.1007/s11214-016-0245-4} {\bibfield  {journal} {\bibinfo  {journal} {Space Sci. Rev.}\ }\textbf {\bibinfo {volume} {199}},\ \bibinfo {pages} {331} (\bibinfo {year} {2016})}\BibitemShut {NoStop}%
\bibitem [{\citenamefont {Gershman}\ \emph {et~al.}(2019)\citenamefont {Gershman}, \citenamefont {Dorelli}, \citenamefont {Avanov}, \citenamefont {Gliese}, \citenamefont {Barrie}, \citenamefont {Schiff}, \citenamefont {Da~Silva}, \citenamefont {Paterson}, \citenamefont {Giles},\ and\ \citenamefont {Pollock}}]{gershman_systematic_2019}%
  \BibitemOpen
  \bibfield  {author} {\bibinfo {author} {\bibfnamefont {D.~J.}\ \bibnamefont {Gershman}}, \bibinfo {author} {\bibfnamefont {J.~C.}\ \bibnamefont {Dorelli}}, \bibinfo {author} {\bibfnamefont {L.~A.}\ \bibnamefont {Avanov}}, \bibinfo {author} {\bibfnamefont {U.}~\bibnamefont {Gliese}}, \bibinfo {author} {\bibfnamefont {A.}~\bibnamefont {Barrie}}, \bibinfo {author} {\bibfnamefont {C.}~\bibnamefont {Schiff}}, \bibinfo {author} {\bibfnamefont {D.~E.}\ \bibnamefont {Da~Silva}}, \bibinfo {author} {\bibfnamefont {W.~R.}\ \bibnamefont {Paterson}}, \bibinfo {author} {\bibfnamefont {B.~L.}\ \bibnamefont {Giles}},\ and\ \bibinfo {author} {\bibfnamefont {C.~J.}\ \bibnamefont {Pollock}},\ }\href {https://doi.org/10.1029/2019JA026980} {\bibfield  {journal} {\bibinfo  {journal} {J. Geophys. Res.}\ }\textbf {\bibinfo {volume} {124}},\ \bibinfo {pages} {10345} (\bibinfo {year} {2019})}\BibitemShut {NoStop}%
\bibitem [{\citenamefont {Svenningsson}\ \emph {et~al.}(2025)\citenamefont {Svenningsson}, \citenamefont {Yordanova}, \citenamefont {Khotyaintsev}, \citenamefont {Andr{\'e}},\ and\ \citenamefont {Cozzani}}]{svenningsson_classifying_2025}%
  \BibitemOpen
  \bibfield  {author} {\bibinfo {author} {\bibfnamefont {I.}~\bibnamefont {Svenningsson}}, \bibinfo {author} {\bibfnamefont {E.}~\bibnamefont {Yordanova}}, \bibinfo {author} {\bibfnamefont {Y.~V.}\ \bibnamefont {Khotyaintsev}}, \bibinfo {author} {\bibfnamefont {M.}~\bibnamefont {Andr{\'e}}},\ and\ \bibinfo {author} {\bibfnamefont {G.}~\bibnamefont {Cozzani}},\ }\href {https://doi.org/10.1029/2024JA033272} {\bibfield  {journal} {\bibinfo  {journal} {J. Geophys. Res.}\ }\textbf {\bibinfo {volume} {130}},\ \bibinfo {pages} {e2024JA033272} (\bibinfo {year} {2025})}\BibitemShut {NoStop}%
\bibitem [{\citenamefont {Retin{\`o}}\ \emph {et~al.}(2007)\citenamefont {Retin{\`o}}, \citenamefont {Sundkvist}, \citenamefont {Vaivads}, \citenamefont {Mozer}, \citenamefont {Andr{\'e}},\ and\ \citenamefont {Owen}}]{retino_situ_2007}%
  \BibitemOpen
  \bibfield  {author} {\bibinfo {author} {\bibfnamefont {A.}~\bibnamefont {Retin{\`o}}}, \bibinfo {author} {\bibfnamefont {D.}~\bibnamefont {Sundkvist}}, \bibinfo {author} {\bibfnamefont {A.}~\bibnamefont {Vaivads}}, \bibinfo {author} {\bibfnamefont {F.}~\bibnamefont {Mozer}}, \bibinfo {author} {\bibfnamefont {M.}~\bibnamefont {Andr{\'e}}},\ and\ \bibinfo {author} {\bibfnamefont {C.~J.}\ \bibnamefont {Owen}},\ }\href {https://doi.org/10.1038/nphys574} {\bibfield  {journal} {\bibinfo  {journal} {Nature Phys}\ }\textbf {\bibinfo {volume} {3}},\ \bibinfo {pages} {235} (\bibinfo {year} {2007})}\BibitemShut {NoStop}%
\bibitem [{\citenamefont {Yordanova}\ \emph {et~al.}(2016)\citenamefont {Yordanova}, \citenamefont {V{\"o}r{\"o}s}, \citenamefont {Varsani}, \citenamefont {Graham}, \citenamefont {Norgren}, \citenamefont {Khotyaintsev}, \citenamefont {Vaivads}, \citenamefont {Eriksson}, \citenamefont {Nakamura}, \citenamefont {Lindqvist}, \citenamefont {Marklund}, \citenamefont {Ergun}, \citenamefont {Magnes}, \citenamefont {Baumjohann}, \citenamefont {Fischer}, \citenamefont {Plaschke}, \citenamefont {Narita}, \citenamefont {Russell}, \citenamefont {Strangeway}, \citenamefont {Le~Contel}, \citenamefont {Pollock}, \citenamefont {Torbert}, \citenamefont {Giles}, \citenamefont {Burch}, \citenamefont {Avanov}, \citenamefont {Dorelli}, \citenamefont {Gershman}, \citenamefont {Paterson}, \citenamefont {Lavraud},\ and\ \citenamefont {Saito}}]{yordanova_electron_2016}%
  \BibitemOpen
  \bibfield  {author} {\bibinfo {author} {\bibfnamefont {E.}~\bibnamefont {Yordanova}}, \bibinfo {author} {\bibfnamefont {Z.}~\bibnamefont {V{\"o}r{\"o}s}}, \bibinfo {author} {\bibfnamefont {A.}~\bibnamefont {Varsani}}, \bibinfo {author} {\bibfnamefont {D.~B.}\ \bibnamefont {Graham}}, \bibinfo {author} {\bibfnamefont {C.}~\bibnamefont {Norgren}}, \bibinfo {author} {\bibfnamefont {{\relax Yu}.~V.}\ \bibnamefont {Khotyaintsev}}, \bibinfo {author} {\bibfnamefont {A.}~\bibnamefont {Vaivads}}, \bibinfo {author} {\bibfnamefont {E.}~\bibnamefont {Eriksson}}, \bibinfo {author} {\bibfnamefont {R.}~\bibnamefont {Nakamura}}, \bibinfo {author} {\bibfnamefont {P.-A.}\ \bibnamefont {Lindqvist}}, \bibinfo {author} {\bibfnamefont {G.}~\bibnamefont {Marklund}}, \bibinfo {author} {\bibfnamefont {R.~E.}\ \bibnamefont {Ergun}}, \bibinfo {author} {\bibfnamefont {W.}~\bibnamefont {Magnes}}, \bibinfo {author} {\bibfnamefont {W.}~\bibnamefont {Baumjohann}}, \bibinfo {author} {\bibfnamefont {D.}~\bibnamefont {Fischer}}, \bibinfo
  {author} {\bibfnamefont {F.}~\bibnamefont {Plaschke}}, \bibinfo {author} {\bibfnamefont {Y.}~\bibnamefont {Narita}}, \bibinfo {author} {\bibfnamefont {C.~T.}\ \bibnamefont {Russell}}, \bibinfo {author} {\bibfnamefont {R.~J.}\ \bibnamefont {Strangeway}}, \bibinfo {author} {\bibfnamefont {O.}~\bibnamefont {Le~Contel}}, \bibinfo {author} {\bibfnamefont {C.}~\bibnamefont {Pollock}}, \bibinfo {author} {\bibfnamefont {R.~B.}\ \bibnamefont {Torbert}}, \bibinfo {author} {\bibfnamefont {B.~J.}\ \bibnamefont {Giles}}, \bibinfo {author} {\bibfnamefont {J.~L.}\ \bibnamefont {Burch}}, \bibinfo {author} {\bibfnamefont {L.~A.}\ \bibnamefont {Avanov}}, \bibinfo {author} {\bibfnamefont {J.~C.}\ \bibnamefont {Dorelli}}, \bibinfo {author} {\bibfnamefont {D.~J.}\ \bibnamefont {Gershman}}, \bibinfo {author} {\bibfnamefont {W.~R.}\ \bibnamefont {Paterson}}, \bibinfo {author} {\bibfnamefont {B.}~\bibnamefont {Lavraud}},\ and\ \bibinfo {author} {\bibfnamefont {Y.}~\bibnamefont {Saito}},\ }\href
  {https://doi.org/10.1002/2016GL069191} {\bibfield  {journal} {\bibinfo  {journal} {Geophys. Res. Lett.}\ }\textbf {\bibinfo {volume} {43}},\ \bibinfo {pages} {5969} (\bibinfo {year} {2016})}\BibitemShut {NoStop}%
\bibitem [{\citenamefont {V{\"o}r{\"o}s}\ \emph {et~al.}(2017)\citenamefont {V{\"o}r{\"o}s}, \citenamefont {Yordanova}, \citenamefont {Varsani}, \citenamefont {Genestreti}, \citenamefont {Khotyaintsev}, \citenamefont {Li}, \citenamefont {Graham}, \citenamefont {Norgren}, \citenamefont {Nakamura}, \citenamefont {Narita}, \citenamefont {Plaschke}, \citenamefont {Magnes}, \citenamefont {Baumjohann}, \citenamefont {Fischer}, \citenamefont {Vaivads}, \citenamefont {Eriksson}, \citenamefont {Lindqvist}, \citenamefont {Marklund}, \citenamefont {Ergun}, \citenamefont {Leitner}, \citenamefont {Leubner}, \citenamefont {Strangeway}, \citenamefont {Le~Contel}, \citenamefont {Pollock}, \citenamefont {Giles}, \citenamefont {Torbert}, \citenamefont {Burch}, \citenamefont {Avanov}, \citenamefont {Dorelli}, \citenamefont {Gershman}, \citenamefont {Paterson}, \citenamefont {Lavraud},\ and\ \citenamefont {Saito}}]{voros_mms_2017}%
  \BibitemOpen
  \bibfield  {author} {\bibinfo {author} {\bibfnamefont {Z.}~\bibnamefont {V{\"o}r{\"o}s}}, \bibinfo {author} {\bibfnamefont {E.}~\bibnamefont {Yordanova}}, \bibinfo {author} {\bibfnamefont {A.}~\bibnamefont {Varsani}}, \bibinfo {author} {\bibfnamefont {K.~J.}\ \bibnamefont {Genestreti}}, \bibinfo {author} {\bibfnamefont {{\relax Yu}.~V.}\ \bibnamefont {Khotyaintsev}}, \bibinfo {author} {\bibfnamefont {W.}~\bibnamefont {Li}}, \bibinfo {author} {\bibfnamefont {D.~B.}\ \bibnamefont {Graham}}, \bibinfo {author} {\bibfnamefont {C.}~\bibnamefont {Norgren}}, \bibinfo {author} {\bibfnamefont {R.}~\bibnamefont {Nakamura}}, \bibinfo {author} {\bibfnamefont {Y.}~\bibnamefont {Narita}}, \bibinfo {author} {\bibfnamefont {F.}~\bibnamefont {Plaschke}}, \bibinfo {author} {\bibfnamefont {W.}~\bibnamefont {Magnes}}, \bibinfo {author} {\bibfnamefont {W.}~\bibnamefont {Baumjohann}}, \bibinfo {author} {\bibfnamefont {D.}~\bibnamefont {Fischer}}, \bibinfo {author} {\bibfnamefont {A.}~\bibnamefont {Vaivads}}, \bibinfo {author}
  {\bibfnamefont {E.}~\bibnamefont {Eriksson}}, \bibinfo {author} {\bibfnamefont {P.-A.}\ \bibnamefont {Lindqvist}}, \bibinfo {author} {\bibfnamefont {G.}~\bibnamefont {Marklund}}, \bibinfo {author} {\bibfnamefont {R.~E.}\ \bibnamefont {Ergun}}, \bibinfo {author} {\bibfnamefont {M.}~\bibnamefont {Leitner}}, \bibinfo {author} {\bibfnamefont {M.~P.}\ \bibnamefont {Leubner}}, \bibinfo {author} {\bibfnamefont {R.~J.}\ \bibnamefont {Strangeway}}, \bibinfo {author} {\bibfnamefont {O.}~\bibnamefont {Le~Contel}}, \bibinfo {author} {\bibfnamefont {C.}~\bibnamefont {Pollock}}, \bibinfo {author} {\bibfnamefont {B.~J.}\ \bibnamefont {Giles}}, \bibinfo {author} {\bibfnamefont {R.~B.}\ \bibnamefont {Torbert}}, \bibinfo {author} {\bibfnamefont {J.~L.}\ \bibnamefont {Burch}}, \bibinfo {author} {\bibfnamefont {L.~A.}\ \bibnamefont {Avanov}}, \bibinfo {author} {\bibfnamefont {J.~C.}\ \bibnamefont {Dorelli}}, \bibinfo {author} {\bibfnamefont {D.~J.}\ \bibnamefont {Gershman}}, \bibinfo {author} {\bibfnamefont {W.~R.}\
  \bibnamefont {Paterson}}, \bibinfo {author} {\bibfnamefont {B.}~\bibnamefont {Lavraud}},\ and\ \bibinfo {author} {\bibfnamefont {Y.}~\bibnamefont {Saito}},\ }\href {https://doi.org/10.1002/2017JA024535} {\bibfield  {journal} {\bibinfo  {journal} {J. Geophys. Res.}\ }\textbf {\bibinfo {volume} {122}},\ \bibinfo {pages} {11,442} (\bibinfo {year} {2017})}\BibitemShut {NoStop}%
\bibitem [{\citenamefont {Stawarz}\ \emph {et~al.}(2022)\citenamefont {Stawarz}, \citenamefont {Eastwood}, \citenamefont {Phan}, \citenamefont {Gingell}, \citenamefont {Pyakurel}, \citenamefont {Shay}, \citenamefont {Robertson}, \citenamefont {Russell},\ and\ \citenamefont {Le~Contel}}]{stawarz_turbulencedriven_2022}%
  \BibitemOpen
  \bibfield  {author} {\bibinfo {author} {\bibfnamefont {J.~E.}\ \bibnamefont {Stawarz}}, \bibinfo {author} {\bibfnamefont {J.~P.}\ \bibnamefont {Eastwood}}, \bibinfo {author} {\bibfnamefont {T.~D.}\ \bibnamefont {Phan}}, \bibinfo {author} {\bibfnamefont {I.~L.}\ \bibnamefont {Gingell}}, \bibinfo {author} {\bibfnamefont {P.~S.}\ \bibnamefont {Pyakurel}}, \bibinfo {author} {\bibfnamefont {M.~A.}\ \bibnamefont {Shay}}, \bibinfo {author} {\bibfnamefont {S.~L.}\ \bibnamefont {Robertson}}, \bibinfo {author} {\bibfnamefont {C.~T.}\ \bibnamefont {Russell}},\ and\ \bibinfo {author} {\bibfnamefont {O.}~\bibnamefont {Le~Contel}},\ }\href {https://doi.org/10.1063/5.0071106} {\bibfield  {journal} {\bibinfo  {journal} {Phys. Plasmas}\ }\textbf {\bibinfo {volume} {29}},\ \bibinfo {pages} {012302} (\bibinfo {year} {2022})}\BibitemShut {NoStop}%
\bibitem [{\citenamefont {Robert}\ \emph {et~al.}(1998)\citenamefont {Robert}, \citenamefont {Dunlop}, \citenamefont {Roux},\ and\ \citenamefont {Chanteur}}]{robert_accuracy_1998}%
  \BibitemOpen
  \bibfield  {author} {\bibinfo {author} {\bibfnamefont {P.}~\bibnamefont {Robert}}, \bibinfo {author} {\bibfnamefont {M.~W.}\ \bibnamefont {Dunlop}}, \bibinfo {author} {\bibfnamefont {A.}~\bibnamefont {Roux}},\ and\ \bibinfo {author} {\bibfnamefont {G.}~\bibnamefont {Chanteur}},\ }in\ \href@noop {} {\emph {\bibinfo {booktitle} {Analysis {{Methods}} for {{Multi-Spacecraft Data}}}}},\ Vol.~\bibinfo {volume} {1},\ \bibinfo {editor} {edited by\ \bibinfo {editor} {\bibfnamefont {G.}~\bibnamefont {Paschmann}}\ and\ \bibinfo {editor} {\bibfnamefont {P.~W.}\ \bibnamefont {Daly}}}\ (\bibinfo {year} {1998})\ pp.\ \bibinfo {pages} {395--418}\BibitemShut {NoStop}%
\bibitem [{\citenamefont {Chanteur}(1998)}]{chanteur_spatial_1998}%
  \BibitemOpen
  \bibfield  {author} {\bibinfo {author} {\bibfnamefont {G.}~\bibnamefont {Chanteur}},\ }in\ \href@noop {} {\emph {\bibinfo {booktitle} {Analysis {{Methods}} for {{Multi-Spacecraft Data}}}}},\ Vol.~\bibinfo {volume} {1}\ (\bibinfo  {publisher} {ISSI Scientific Reports Series},\ \bibinfo {year} {1998})\ pp.\ \bibinfo {pages} {349--370}\BibitemShut {NoStop}%
\bibitem [{\citenamefont {Chanteur}\ and\ \citenamefont {Harvey}(1998)}]{chanteur_spatial_1998a}%
  \BibitemOpen
  \bibfield  {author} {\bibinfo {author} {\bibfnamefont {G.}~\bibnamefont {Chanteur}}\ and\ \bibinfo {author} {\bibfnamefont {C.~C.}\ \bibnamefont {Harvey}},\ }in\ \href@noop {} {\emph {\bibinfo {booktitle} {Analysis {{Methods}} for {{Multi-Spacecraft Data}}}}},\ Vol.~\bibinfo {volume} {1}\ (\bibinfo  {publisher} {ISSI Scientific Reports Series},\ \bibinfo {year} {1998})\ pp.\ \bibinfo {pages} {371--394}\BibitemShut {NoStop}%
\bibitem [{\citenamefont {Swisdak}(2016)}]{swisdak_quantifying_2016}%
  \BibitemOpen
  \bibfield  {author} {\bibinfo {author} {\bibfnamefont {M.}~\bibnamefont {Swisdak}},\ }\href {https://doi.org/10.1002/2015GL066980} {\bibfield  {journal} {\bibinfo  {journal} {Geophys. Res. Lett.}\ }\textbf {\bibinfo {volume} {43}},\ \bibinfo {pages} {43} (\bibinfo {year} {2016})}\BibitemShut {NoStop}%
\bibitem [{\citenamefont {Verscharen}\ \emph {et~al.}(2016)\citenamefont {Verscharen}, \citenamefont {Chandran}, \citenamefont {Klein},\ and\ \citenamefont {Quataert}}]{verscharen_collisionless_2016}%
  \BibitemOpen
  \bibfield  {author} {\bibinfo {author} {\bibfnamefont {D.}~\bibnamefont {Verscharen}}, \bibinfo {author} {\bibfnamefont {B.~D.~G.}\ \bibnamefont {Chandran}}, \bibinfo {author} {\bibfnamefont {K.~G.}\ \bibnamefont {Klein}},\ and\ \bibinfo {author} {\bibfnamefont {E.}~\bibnamefont {Quataert}},\ }\href {https://doi.org/10.3847/0004-637X/831/2/128} {\bibfield  {journal} {\bibinfo  {journal} {Astrophys. J.}\ }\textbf {\bibinfo {volume} {831}},\ \bibinfo {pages} {128} (\bibinfo {year} {2016})}\BibitemShut {NoStop}%
\bibitem [{\citenamefont {Barlow}(1989)}]{barlow_statistics_1989}%
  \BibitemOpen
  \bibfield  {author} {\bibinfo {author} {\bibfnamefont {R.}~\bibnamefont {Barlow}},\ }\href@noop {} {\emph {\bibinfo {title} {Statistics. {{A}} Guide to the Use of Statistical Methods in the Physical Sciences}}}\ (\bibinfo  {publisher} {The Manchester Physics Series},\ \bibinfo {year} {1989})\BibitemShut {NoStop}%
\bibitem [{\citenamefont {Motoba}\ \emph {et~al.}(2022)\citenamefont {Motoba}, \citenamefont {Sitnov}, \citenamefont {Stephens},\ and\ \citenamefont {Gershman}}]{motoba_new_2022}%
  \BibitemOpen
  \bibfield  {author} {\bibinfo {author} {\bibfnamefont {T.}~\bibnamefont {Motoba}}, \bibinfo {author} {\bibfnamefont {M.~I.}\ \bibnamefont {Sitnov}}, \bibinfo {author} {\bibfnamefont {G.~K.}\ \bibnamefont {Stephens}},\ and\ \bibinfo {author} {\bibfnamefont {D.~J.}\ \bibnamefont {Gershman}},\ }\href {https://doi.org/10.1029/2022JA030514} {\bibfield  {journal} {\bibinfo  {journal} {J. Geophys. Res.}\ }\textbf {\bibinfo {volume} {127}},\ \bibinfo {pages} {e2022JA030514} (\bibinfo {year} {2022})}\BibitemShut {NoStop}%
\bibitem [{\citenamefont {Graham}\ \emph {et~al.}(2021)\citenamefont {Graham}, \citenamefont {Khotyaintsev}, \citenamefont {Andr{\'e}}, \citenamefont {Vaivads}, \citenamefont {Chasapis}, \citenamefont {Matthaeus}, \citenamefont {Retin{\`o}}, \citenamefont {Valentini},\ and\ \citenamefont {Gershman}}]{graham_nonmaxwellianity_2021}%
  \BibitemOpen
  \bibfield  {author} {\bibinfo {author} {\bibfnamefont {D.~B.}\ \bibnamefont {Graham}}, \bibinfo {author} {\bibfnamefont {Y.~V.}\ \bibnamefont {Khotyaintsev}}, \bibinfo {author} {\bibfnamefont {M.}~\bibnamefont {Andr{\'e}}}, \bibinfo {author} {\bibfnamefont {A.}~\bibnamefont {Vaivads}}, \bibinfo {author} {\bibfnamefont {A.}~\bibnamefont {Chasapis}}, \bibinfo {author} {\bibfnamefont {W.~H.}\ \bibnamefont {Matthaeus}}, \bibinfo {author} {\bibfnamefont {A.}~\bibnamefont {Retin{\`o}}}, \bibinfo {author} {\bibfnamefont {F.}~\bibnamefont {Valentini}},\ and\ \bibinfo {author} {\bibfnamefont {D.~J.}\ \bibnamefont {Gershman}},\ }\href {https://doi.org/10.1029/2021JA029260} {\bibfield  {journal} {\bibinfo  {journal} {J. Geophys. Res.}\ }\textbf {\bibinfo {volume} {126}},\ \bibinfo {pages} {e2021JA029260} (\bibinfo {year} {2021})}\BibitemShut {NoStop}%
\bibitem [{\citenamefont {Servidio}\ \emph {et~al.}(2015)\citenamefont {Servidio}, \citenamefont {Valentini}, \citenamefont {Perrone}, \citenamefont {Greco}, \citenamefont {Califano}, \citenamefont {Matthaeus},\ and\ \citenamefont {Veltri}}]{servidio_kinetic_2015}%
  \BibitemOpen
  \bibfield  {author} {\bibinfo {author} {\bibfnamefont {S.}~\bibnamefont {Servidio}}, \bibinfo {author} {\bibfnamefont {F.}~\bibnamefont {Valentini}}, \bibinfo {author} {\bibfnamefont {D.}~\bibnamefont {Perrone}}, \bibinfo {author} {\bibfnamefont {A.}~\bibnamefont {Greco}}, \bibinfo {author} {\bibfnamefont {F.}~\bibnamefont {Califano}}, \bibinfo {author} {\bibfnamefont {W.~H.}\ \bibnamefont {Matthaeus}},\ and\ \bibinfo {author} {\bibfnamefont {P.}~\bibnamefont {Veltri}},\ }\href {https://doi.org/10.1017/S0022377814000841} {\bibfield  {journal} {\bibinfo  {journal} {J. Plasma Phys.}\ }\textbf {\bibinfo {volume} {81}},\ \bibinfo {pages} {325810107} (\bibinfo {year} {2015})}\BibitemShut {NoStop}%
\bibitem [{\citenamefont {Servidio}\ \emph {et~al.}(2014)\citenamefont {Servidio}, \citenamefont {Osman}, \citenamefont {Valentini}, \citenamefont {Perrone}, \citenamefont {Califano}, \citenamefont {Chapman}, \citenamefont {Matthaeus},\ and\ \citenamefont {Veltri}}]{servidio_proton_2014}%
  \BibitemOpen
  \bibfield  {author} {\bibinfo {author} {\bibfnamefont {S.}~\bibnamefont {Servidio}}, \bibinfo {author} {\bibfnamefont {K.~T.}\ \bibnamefont {Osman}}, \bibinfo {author} {\bibfnamefont {F.}~\bibnamefont {Valentini}}, \bibinfo {author} {\bibfnamefont {D.}~\bibnamefont {Perrone}}, \bibinfo {author} {\bibfnamefont {F.}~\bibnamefont {Califano}}, \bibinfo {author} {\bibfnamefont {S.}~\bibnamefont {Chapman}}, \bibinfo {author} {\bibfnamefont {W.~H.}\ \bibnamefont {Matthaeus}},\ and\ \bibinfo {author} {\bibfnamefont {P.}~\bibnamefont {Veltri}},\ }\href {https://doi.org/10.1088/2041-8205/781/2/L27} {\bibfield  {journal} {\bibinfo  {journal} {Astrophys. J. Lett.}\ }\textbf {\bibinfo {volume} {781}},\ \bibinfo {pages} {L27} (\bibinfo {year} {2014})}\BibitemShut {NoStop}%
\bibitem [{\citenamefont {B{\"u}chner}\ and\ \citenamefont {Zelenyi}(1989)}]{buchner_regular_1989}%
  \BibitemOpen
  \bibfield  {author} {\bibinfo {author} {\bibfnamefont {J.}~\bibnamefont {B{\"u}chner}}\ and\ \bibinfo {author} {\bibfnamefont {L.~M.}\ \bibnamefont {Zelenyi}},\ }\href {https://doi.org/10.1029/JA094iA09p11821} {\bibfield  {journal} {\bibinfo  {journal} {J. Geophys. Res.}\ }\textbf {\bibinfo {volume} {94}},\ \bibinfo {pages} {11821} (\bibinfo {year} {1989})}\BibitemShut {NoStop}%
\bibitem [{\citenamefont {Yang}\ \emph {et~al.}(2019)\citenamefont {Yang}, \citenamefont {Wan}, \citenamefont {Matthaeus}, \citenamefont {Shi}, \citenamefont {Parashar}, \citenamefont {Lu},\ and\ \citenamefont {Chen}}]{yang_role_2019}%
  \BibitemOpen
  \bibfield  {author} {\bibinfo {author} {\bibfnamefont {Y.}~\bibnamefont {Yang}}, \bibinfo {author} {\bibfnamefont {M.}~\bibnamefont {Wan}}, \bibinfo {author} {\bibfnamefont {W.~H.}\ \bibnamefont {Matthaeus}}, \bibinfo {author} {\bibfnamefont {Y.}~\bibnamefont {Shi}}, \bibinfo {author} {\bibfnamefont {T.~N.}\ \bibnamefont {Parashar}}, \bibinfo {author} {\bibfnamefont {Q.}~\bibnamefont {Lu}},\ and\ \bibinfo {author} {\bibfnamefont {S.}~\bibnamefont {Chen}},\ }\href {https://doi.org/10.1063/1.5099360} {\bibfield  {journal} {\bibinfo  {journal} {Phys. Plasmas}\ }\textbf {\bibinfo {volume} {26}},\ \bibinfo {pages} {072306} (\bibinfo {year} {2019})}\BibitemShut {NoStop}%
\bibitem [{\citenamefont {Bandyopadhyay}\ \emph {et~al.}(2020)\citenamefont {Bandyopadhyay}, \citenamefont {Yang}, \citenamefont {Matthaeus}, \citenamefont {Chasapis}, \citenamefont {Parashar}, \citenamefont {Russell}, \citenamefont {Strangeway}, \citenamefont {Torbert}, \citenamefont {Giles}, \citenamefont {Gershman}, \citenamefont {Pollock}, \citenamefont {Moore},\ and\ \citenamefont {Burch}}]{bandyopadhyay_situ_2020a}%
  \BibitemOpen
  \bibfield  {author} {\bibinfo {author} {\bibfnamefont {R.}~\bibnamefont {Bandyopadhyay}}, \bibinfo {author} {\bibfnamefont {Y.}~\bibnamefont {Yang}}, \bibinfo {author} {\bibfnamefont {W.~H.}\ \bibnamefont {Matthaeus}}, \bibinfo {author} {\bibfnamefont {A.}~\bibnamefont {Chasapis}}, \bibinfo {author} {\bibfnamefont {T.~N.}\ \bibnamefont {Parashar}}, \bibinfo {author} {\bibfnamefont {C.~T.}\ \bibnamefont {Russell}}, \bibinfo {author} {\bibfnamefont {R.~J.}\ \bibnamefont {Strangeway}}, \bibinfo {author} {\bibfnamefont {R.~B.}\ \bibnamefont {Torbert}}, \bibinfo {author} {\bibfnamefont {B.~L.}\ \bibnamefont {Giles}}, \bibinfo {author} {\bibfnamefont {D.~J.}\ \bibnamefont {Gershman}}, \bibinfo {author} {\bibfnamefont {C.~J.}\ \bibnamefont {Pollock}}, \bibinfo {author} {\bibfnamefont {T.~E.}\ \bibnamefont {Moore}},\ and\ \bibinfo {author} {\bibfnamefont {J.~L.}\ \bibnamefont {Burch}},\ }\href {https://doi.org/10.3847/2041-8213/ab846e} {\bibfield  {journal} {\bibinfo  {journal} {Astrophys. J. Lett.}\ }\textbf
  {\bibinfo {volume} {893}},\ \bibinfo {pages} {L25} (\bibinfo {year} {2020})}\BibitemShut {NoStop}%
\bibitem [{\citenamefont {Birmingham}(1984)}]{birmingham_pitch_1984}%
  \BibitemOpen
  \bibfield  {author} {\bibinfo {author} {\bibfnamefont {T.~J.}\ \bibnamefont {Birmingham}},\ }\href {https://doi.org/10.1029/JA089iA05p02699} {\bibfield  {journal} {\bibinfo  {journal} {J. Geophys. Res.}\ }\textbf {\bibinfo {volume} {89}},\ \bibinfo {pages} {2699} (\bibinfo {year} {1984})}\BibitemShut {NoStop}%
\bibitem [{\citenamefont {Burkhart}\ \emph {et~al.}(1995)\citenamefont {Burkhart}, \citenamefont {Dusenbery},\ and\ \citenamefont {Speiser}}]{burkhart_particle_1995}%
  \BibitemOpen
  \bibfield  {author} {\bibinfo {author} {\bibfnamefont {G.~R.}\ \bibnamefont {Burkhart}}, \bibinfo {author} {\bibfnamefont {P.~B.}\ \bibnamefont {Dusenbery}},\ and\ \bibinfo {author} {\bibfnamefont {T.~W.}\ \bibnamefont {Speiser}},\ }\href {https://doi.org/10.1029/94JA02221} {\bibfield  {journal} {\bibinfo  {journal} {J. Geophys. Res.}\ }\textbf {\bibinfo {volume} {100}},\ \bibinfo {pages} {107} (\bibinfo {year} {1995})}\BibitemShut {NoStop}%
\bibitem [{\citenamefont {Artemyev}\ \emph {et~al.}(2014)\citenamefont {Artemyev}, \citenamefont {Neishtadt},\ and\ \citenamefont {Zelenyi}}]{artemyev_rapid_2014}%
  \BibitemOpen
  \bibfield  {author} {\bibinfo {author} {\bibfnamefont {A.~V.}\ \bibnamefont {Artemyev}}, \bibinfo {author} {\bibfnamefont {A.~I.}\ \bibnamefont {Neishtadt}},\ and\ \bibinfo {author} {\bibfnamefont {L.~M.}\ \bibnamefont {Zelenyi}},\ }\href {https://doi.org/10.1103/PhysRevE.89.060902} {\bibfield  {journal} {\bibinfo  {journal} {Phys. Rev. E}\ }\textbf {\bibinfo {volume} {89}},\ \bibinfo {pages} {060902} (\bibinfo {year} {2014})}\BibitemShut {NoStop}%
\bibitem [{\citenamefont {Klein}\ \emph {et~al.}(2018)\citenamefont {Klein}, \citenamefont {Alterman}, \citenamefont {Stevens}, \citenamefont {Vech},\ and\ \citenamefont {Kasper}}]{klein_majority_2018}%
  \BibitemOpen
  \bibfield  {author} {\bibinfo {author} {\bibfnamefont {K.~G.}\ \bibnamefont {Klein}}, \bibinfo {author} {\bibfnamefont {B.~L.}\ \bibnamefont {Alterman}}, \bibinfo {author} {\bibfnamefont {M.~L.}\ \bibnamefont {Stevens}}, \bibinfo {author} {\bibfnamefont {D.}~\bibnamefont {Vech}},\ and\ \bibinfo {author} {\bibfnamefont {J.~C.}\ \bibnamefont {Kasper}},\ }\href {https://doi.org/10.1103/PhysRevLett.120.205102} {\bibfield  {journal} {\bibinfo  {journal} {Phys. Rev. Lett.}\ }\textbf {\bibinfo {volume} {120}},\ \bibinfo {pages} {205102} (\bibinfo {year} {2018})}\BibitemShut {NoStop}%
\bibitem [{\citenamefont {Martinovi{\'c}}\ \emph {et~al.}(2021)\citenamefont {Martinovi{\'c}}, \citenamefont {Klein}, \citenamefont {{\v D}urovcov{\'a}},\ and\ \citenamefont {Alterman}}]{martinovic_iondriven_2021}%
  \BibitemOpen
  \bibfield  {author} {\bibinfo {author} {\bibfnamefont {M.~M.}\ \bibnamefont {Martinovi{\'c}}}, \bibinfo {author} {\bibfnamefont {K.~G.}\ \bibnamefont {Klein}}, \bibinfo {author} {\bibfnamefont {T.}~\bibnamefont {{\v D}urovcov{\'a}}},\ and\ \bibinfo {author} {\bibfnamefont {B.~L.}\ \bibnamefont {Alterman}},\ }\href {https://doi.org/10.3847/1538-4357/ac3081} {\bibfield  {journal} {\bibinfo  {journal} {Astrophys. J.}\ }\textbf {\bibinfo {volume} {923}},\ \bibinfo {pages} {116} (\bibinfo {year} {2021})}\BibitemShut {NoStop}%
\bibitem [{\citenamefont {Martinovi{\'c}}\ and\ \citenamefont {Klein}(2023)}]{martinovic_iondriven_2023}%
  \BibitemOpen
  \bibfield  {author} {\bibinfo {author} {\bibfnamefont {M.~M.}\ \bibnamefont {Martinovi{\'c}}}\ and\ \bibinfo {author} {\bibfnamefont {K.~G.}\ \bibnamefont {Klein}},\ }\href {https://doi.org/10.3847/1538-4357/acdb79} {\bibfield  {journal} {\bibinfo  {journal} {Astrophys. J.}\ }\textbf {\bibinfo {volume} {952}},\ \bibinfo {pages} {14} (\bibinfo {year} {2023})}\BibitemShut {NoStop}%
\bibitem [{\citenamefont {Maruca}\ \emph {et~al.}(2012)\citenamefont {Maruca}, \citenamefont {Kasper},\ and\ \citenamefont {Gary}}]{maruca_instabilitydriven_2012}%
  \BibitemOpen
  \bibfield  {author} {\bibinfo {author} {\bibfnamefont {B.~A.}\ \bibnamefont {Maruca}}, \bibinfo {author} {\bibfnamefont {J.~C.}\ \bibnamefont {Kasper}},\ and\ \bibinfo {author} {\bibfnamefont {S.~P.}\ \bibnamefont {Gary}},\ }\href {https://doi.org/10.1088/0004-637X/748/2/137} {\bibfield  {journal} {\bibinfo  {journal} {ApJ}\ }\textbf {\bibinfo {volume} {748}},\ \bibinfo {pages} {137} (\bibinfo {year} {2012})}\BibitemShut {NoStop}%
\bibitem [{\citenamefont {Chen}\ \emph {et~al.}(2016)\citenamefont {Chen}, \citenamefont {Matteini}, \citenamefont {Schekochihin}, \citenamefont {Stevens}, \citenamefont {Salem}, \citenamefont {Maruca}, \citenamefont {Kunz},\ and\ \citenamefont {Bale}}]{chen_multispecies_2016}%
  \BibitemOpen
  \bibfield  {author} {\bibinfo {author} {\bibfnamefont {C.~H.~K.}\ \bibnamefont {Chen}}, \bibinfo {author} {\bibfnamefont {L.}~\bibnamefont {Matteini}}, \bibinfo {author} {\bibfnamefont {A.~A.}\ \bibnamefont {Schekochihin}}, \bibinfo {author} {\bibfnamefont {M.~L.}\ \bibnamefont {Stevens}}, \bibinfo {author} {\bibfnamefont {C.~S.}\ \bibnamefont {Salem}}, \bibinfo {author} {\bibfnamefont {B.~A.}\ \bibnamefont {Maruca}}, \bibinfo {author} {\bibfnamefont {M.~W.}\ \bibnamefont {Kunz}},\ and\ \bibinfo {author} {\bibfnamefont {S.~D.}\ \bibnamefont {Bale}},\ }\href {https://doi.org/10.3847/2041-8205/825/2/L26} {\bibfield  {journal} {\bibinfo  {journal} {ApJL}\ }\textbf {\bibinfo {volume} {825}},\ \bibinfo {pages} {L26} (\bibinfo {year} {2016})}\BibitemShut {NoStop}%
\bibitem [{\citenamefont {DeWeese}\ \emph {et~al.}(2022)\citenamefont {DeWeese}, \citenamefont {Maruca}, \citenamefont {Qudsi}, \citenamefont {Chasapis}, \citenamefont {Pultrone}, \citenamefont {Johnson}, \citenamefont {Vines}, \citenamefont {Shay}, \citenamefont {Matthaeus}, \citenamefont {Gomez}, \citenamefont {Fuselier}, \citenamefont {Giles}, \citenamefont {Gershman}, \citenamefont {Russell}, \citenamefont {Strangeway}, \citenamefont {Burch},\ and\ \citenamefont {Torbert}}]{deweese_alpha_2022}%
  \BibitemOpen
  \bibfield  {author} {\bibinfo {author} {\bibfnamefont {H.}~\bibnamefont {DeWeese}}, \bibinfo {author} {\bibfnamefont {B.~A.}\ \bibnamefont {Maruca}}, \bibinfo {author} {\bibfnamefont {R.~A.}\ \bibnamefont {Qudsi}}, \bibinfo {author} {\bibfnamefont {A.}~\bibnamefont {Chasapis}}, \bibinfo {author} {\bibfnamefont {M.}~\bibnamefont {Pultrone}}, \bibinfo {author} {\bibfnamefont {E.}~\bibnamefont {Johnson}}, \bibinfo {author} {\bibfnamefont {S.~K.}\ \bibnamefont {Vines}}, \bibinfo {author} {\bibfnamefont {M.~A.}\ \bibnamefont {Shay}}, \bibinfo {author} {\bibfnamefont {W.~H.}\ \bibnamefont {Matthaeus}}, \bibinfo {author} {\bibfnamefont {R.~G.}\ \bibnamefont {Gomez}}, \bibinfo {author} {\bibfnamefont {S.~A.}\ \bibnamefont {Fuselier}}, \bibinfo {author} {\bibfnamefont {B.~L.}\ \bibnamefont {Giles}}, \bibinfo {author} {\bibfnamefont {D.~J.}\ \bibnamefont {Gershman}}, \bibinfo {author} {\bibfnamefont {C.~T.}\ \bibnamefont {Russell}}, \bibinfo {author} {\bibfnamefont {R.~J.}\ \bibnamefont {Strangeway}}, \bibinfo
  {author} {\bibfnamefont {J.~L.}\ \bibnamefont {Burch}},\ and\ \bibinfo {author} {\bibfnamefont {R.~B.}\ \bibnamefont {Torbert}},\ }\href {https://doi.org/10.3847/1538-4357/ac9791} {\bibfield  {journal} {\bibinfo  {journal} {ApJ}\ }\textbf {\bibinfo {volume} {941}},\ \bibinfo {pages} {12} (\bibinfo {year} {2022})}\BibitemShut {NoStop}%
\bibitem [{\citenamefont {Osman}\ \emph {et~al.}(2012)\citenamefont {Osman}, \citenamefont {Matthaeus}, \citenamefont {Hnat},\ and\ \citenamefont {Chapman}}]{osman_kinetic_2012}%
  \BibitemOpen
  \bibfield  {author} {\bibinfo {author} {\bibfnamefont {K.~T.}\ \bibnamefont {Osman}}, \bibinfo {author} {\bibfnamefont {W.~H.}\ \bibnamefont {Matthaeus}}, \bibinfo {author} {\bibfnamefont {B.}~\bibnamefont {Hnat}},\ and\ \bibinfo {author} {\bibfnamefont {S.~C.}\ \bibnamefont {Chapman}},\ }\href {https://doi.org/10.1103/PhysRevLett.108.261103} {\bibfield  {journal} {\bibinfo  {journal} {Phys. Rev. Lett.}\ }\textbf {\bibinfo {volume} {108}},\ \bibinfo {pages} {261103} (\bibinfo {year} {2012})}\BibitemShut {NoStop}%
\bibitem [{\citenamefont {Kunz}\ \emph {et~al.}(2014)\citenamefont {Kunz}, \citenamefont {Schekochihin},\ and\ \citenamefont {Stone}}]{kunz_firehose_2014}%
  \BibitemOpen
  \bibfield  {author} {\bibinfo {author} {\bibfnamefont {M.~W.}\ \bibnamefont {Kunz}}, \bibinfo {author} {\bibfnamefont {A.~A.}\ \bibnamefont {Schekochihin}},\ and\ \bibinfo {author} {\bibfnamefont {J.~M.}\ \bibnamefont {Stone}},\ }\href {https://doi.org/10.1103/PhysRevLett.112.205003} {\bibfield  {journal} {\bibinfo  {journal} {Phys. Rev. Lett.}\ }\textbf {\bibinfo {volume} {112}},\ \bibinfo {pages} {205003} (\bibinfo {year} {2014})}\BibitemShut {NoStop}%
\bibitem [{\citenamefont {Del~Sarto}\ \emph {et~al.}(2016)\citenamefont {Del~Sarto}, \citenamefont {Pegoraro},\ and\ \citenamefont {Califano}}]{delsarto_pressure_2016}%
  \BibitemOpen
  \bibfield  {author} {\bibinfo {author} {\bibfnamefont {D.}~\bibnamefont {Del~Sarto}}, \bibinfo {author} {\bibfnamefont {F.}~\bibnamefont {Pegoraro}},\ and\ \bibinfo {author} {\bibfnamefont {F.}~\bibnamefont {Califano}},\ }\href {https://doi.org/10.1103/PhysRevE.93.053203} {\bibfield  {journal} {\bibinfo  {journal} {Phys. Rev. E}\ }\textbf {\bibinfo {volume} {93}},\ \bibinfo {pages} {053203} (\bibinfo {year} {2016})}\BibitemShut {NoStop}%
\bibitem [{\citenamefont {Sundkvist}\ \emph {et~al.}(2007)\citenamefont {Sundkvist}, \citenamefont {Retin{\`o}}, \citenamefont {Vaivads},\ and\ \citenamefont {Bale}}]{sundkvist_dissipation_2007}%
  \BibitemOpen
  \bibfield  {author} {\bibinfo {author} {\bibfnamefont {D.}~\bibnamefont {Sundkvist}}, \bibinfo {author} {\bibfnamefont {A.}~\bibnamefont {Retin{\`o}}}, \bibinfo {author} {\bibfnamefont {A.}~\bibnamefont {Vaivads}},\ and\ \bibinfo {author} {\bibfnamefont {S.~D.}\ \bibnamefont {Bale}},\ }\href {https://doi.org/10.1103/PhysRevLett.99.025004} {\bibfield  {journal} {\bibinfo  {journal} {Phys. Rev. Lett.}\ }\textbf {\bibinfo {volume} {99}},\ \bibinfo {pages} {025004} (\bibinfo {year} {2007})}\BibitemShut {NoStop}%
\bibitem [{\citenamefont {Yordanova}\ \emph {et~al.}(2008)\citenamefont {Yordanova}, \citenamefont {Vaivads}, \citenamefont {Andr{\'e}}, \citenamefont {Buchert},\ and\ \citenamefont {V{\"o}r{\"o}s}}]{yordanova_magnetosheath_2008}%
  \BibitemOpen
  \bibfield  {author} {\bibinfo {author} {\bibfnamefont {E.}~\bibnamefont {Yordanova}}, \bibinfo {author} {\bibfnamefont {A.}~\bibnamefont {Vaivads}}, \bibinfo {author} {\bibfnamefont {M.}~\bibnamefont {Andr{\'e}}}, \bibinfo {author} {\bibfnamefont {S.~C.}\ \bibnamefont {Buchert}},\ and\ \bibinfo {author} {\bibfnamefont {Z.}~\bibnamefont {V{\"o}r{\"o}s}},\ }\href {https://doi.org/10.1103/PhysRevLett.100.205003} {\bibfield  {journal} {\bibinfo  {journal} {Phys. Rev. Lett.}\ }\textbf {\bibinfo {volume} {100}},\ \bibinfo {pages} {205003} (\bibinfo {year} {2008})}\BibitemShut {NoStop}%
\bibitem [{\citenamefont {Tatsuno}\ \emph {et~al.}(2009)\citenamefont {Tatsuno}, \citenamefont {Dorland}, \citenamefont {Schekochihin}, \citenamefont {Plunk}, \citenamefont {Barnes}, \citenamefont {Cowley},\ and\ \citenamefont {Howes}}]{tatsuno_nonlinear_2009}%
  \BibitemOpen
  \bibfield  {author} {\bibinfo {author} {\bibfnamefont {T.}~\bibnamefont {Tatsuno}}, \bibinfo {author} {\bibfnamefont {W.}~\bibnamefont {Dorland}}, \bibinfo {author} {\bibfnamefont {A.~A.}\ \bibnamefont {Schekochihin}}, \bibinfo {author} {\bibfnamefont {G.~G.}\ \bibnamefont {Plunk}}, \bibinfo {author} {\bibfnamefont {M.}~\bibnamefont {Barnes}}, \bibinfo {author} {\bibfnamefont {S.~C.}\ \bibnamefont {Cowley}},\ and\ \bibinfo {author} {\bibfnamefont {G.~G.}\ \bibnamefont {Howes}},\ }\href {https://doi.org/10.1103/PhysRevLett.103.015003} {\bibfield  {journal} {\bibinfo  {journal} {Phys. Rev. Lett.}\ }\textbf {\bibinfo {volume} {103}},\ \bibinfo {pages} {015003} (\bibinfo {year} {2009})}\BibitemShut {NoStop}%
\bibitem [{\citenamefont {Servidio}\ \emph {et~al.}(2017)\citenamefont {Servidio}, \citenamefont {Chasapis}, \citenamefont {Matthaeus}, \citenamefont {Perrone}, \citenamefont {Valentini}, \citenamefont {Parashar}, \citenamefont {Veltri}, \citenamefont {Gershman}, \citenamefont {Russell}, \citenamefont {Giles}, \citenamefont {Fuselier}, \citenamefont {Phan},\ and\ \citenamefont {Burch}}]{servidio_magnetospheric_2017}%
  \BibitemOpen
  \bibfield  {author} {\bibinfo {author} {\bibfnamefont {S.}~\bibnamefont {Servidio}}, \bibinfo {author} {\bibfnamefont {A.}~\bibnamefont {Chasapis}}, \bibinfo {author} {\bibfnamefont {W.~H.}\ \bibnamefont {Matthaeus}}, \bibinfo {author} {\bibfnamefont {D.}~\bibnamefont {Perrone}}, \bibinfo {author} {\bibfnamefont {F.}~\bibnamefont {Valentini}}, \bibinfo {author} {\bibfnamefont {T.~N.}\ \bibnamefont {Parashar}}, \bibinfo {author} {\bibfnamefont {P.}~\bibnamefont {Veltri}}, \bibinfo {author} {\bibfnamefont {D.}~\bibnamefont {Gershman}}, \bibinfo {author} {\bibfnamefont {C.~T.}\ \bibnamefont {Russell}}, \bibinfo {author} {\bibfnamefont {B.}~\bibnamefont {Giles}}, \bibinfo {author} {\bibfnamefont {S.~A.}\ \bibnamefont {Fuselier}}, \bibinfo {author} {\bibfnamefont {T.~D.}\ \bibnamefont {Phan}},\ and\ \bibinfo {author} {\bibfnamefont {J.}~\bibnamefont {Burch}},\ }\href {https://doi.org/10.1103/PhysRevLett.119.205101} {\bibfield  {journal} {\bibinfo  {journal} {Phys. Rev. Lett.}\ }\textbf {\bibinfo {volume}
  {119}},\ \bibinfo {pages} {205101} (\bibinfo {year} {2017})}\BibitemShut {NoStop}%
\bibitem [{\citenamefont {Artemyev}\ \emph {et~al.}(2016)\citenamefont {Artemyev}, \citenamefont {Vainchtein}, \citenamefont {Neishtadt},\ and\ \citenamefont {Zelenyi}}]{artemyev_charged_2016}%
  \BibitemOpen
  \bibfield  {author} {\bibinfo {author} {\bibfnamefont {A.~V.}\ \bibnamefont {Artemyev}}, \bibinfo {author} {\bibfnamefont {D.~L.}\ \bibnamefont {Vainchtein}}, \bibinfo {author} {\bibfnamefont {A.~I.}\ \bibnamefont {Neishtadt}},\ and\ \bibinfo {author} {\bibfnamefont {L.~M.}\ \bibnamefont {Zelenyi}},\ }\href {https://doi.org/10.1103/PhysRevE.93.053207} {\bibfield  {journal} {\bibinfo  {journal} {Phys. Rev. E}\ }\textbf {\bibinfo {volume} {93}},\ \bibinfo {pages} {053207} (\bibinfo {year} {2016})}\BibitemShut {NoStop}%
\bibitem [{\citenamefont {Malara}\ \emph {et~al.}(2021)\citenamefont {Malara}, \citenamefont {Perri},\ and\ \citenamefont {Zimbardo}}]{malara_chargedparticle_2021}%
  \BibitemOpen
  \bibfield  {author} {\bibinfo {author} {\bibfnamefont {F.}~\bibnamefont {Malara}}, \bibinfo {author} {\bibfnamefont {S.}~\bibnamefont {Perri}},\ and\ \bibinfo {author} {\bibfnamefont {G.}~\bibnamefont {Zimbardo}},\ }\href {https://doi.org/10.1103/PhysRevE.104.025208} {\bibfield  {journal} {\bibinfo  {journal} {Phys. Rev. E}\ }\textbf {\bibinfo {volume} {104}},\ \bibinfo {pages} {025208} (\bibinfo {year} {2021})}\BibitemShut {NoStop}%
\bibitem [{\citenamefont {Servidio}\ \emph {et~al.}(2016)\citenamefont {Servidio}, \citenamefont {Haynes}, \citenamefont {Matthaeus}, \citenamefont {Burgess}, \citenamefont {Carbone},\ and\ \citenamefont {Veltri}}]{servidio_explosive_2016}%
  \BibitemOpen
  \bibfield  {author} {\bibinfo {author} {\bibfnamefont {S.}~\bibnamefont {Servidio}}, \bibinfo {author} {\bibfnamefont {C.~T.}\ \bibnamefont {Haynes}}, \bibinfo {author} {\bibfnamefont {W.~H.}\ \bibnamefont {Matthaeus}}, \bibinfo {author} {\bibfnamefont {D.}~\bibnamefont {Burgess}}, \bibinfo {author} {\bibfnamefont {V.}~\bibnamefont {Carbone}},\ and\ \bibinfo {author} {\bibfnamefont {P.}~\bibnamefont {Veltri}},\ }\href {https://doi.org/10.1103/PhysRevLett.117.095101} {\bibfield  {journal} {\bibinfo  {journal} {Phys. Rev. Lett.}\ }\textbf {\bibinfo {volume} {117}},\ \bibinfo {pages} {095101} (\bibinfo {year} {2016})}\BibitemShut {NoStop}%
\bibitem [{\citenamefont {Lemoine}(2022)}]{lemoine_firstprinciples_2022}%
  \BibitemOpen
  \bibfield  {author} {\bibinfo {author} {\bibfnamefont {M.}~\bibnamefont {Lemoine}},\ }\href {https://doi.org/10.1103/PhysRevLett.129.215101} {\bibfield  {journal} {\bibinfo  {journal} {Phys. Rev. Lett.}\ }\textbf {\bibinfo {volume} {129}},\ \bibinfo {pages} {215101} (\bibinfo {year} {2022})}\BibitemShut {NoStop}%
\bibitem [{Note1()}]{Note1}%
  \BibitemOpen
  \bibinfo {note} {See \protect \url {https://pypi.org/project/pyrfu/}.}\BibitemShut {Stop}%
\end{thebibliography}%

\end{document}